\documentclass{article}

\usepackage{arxiv}

\usepackage[utf8]{inputenc} 
\usepackage[T1]{fontenc}    
\usepackage{hyperref}       
\usepackage{url}            
\usepackage{booktabs}       
\usepackage{amsfonts}       
\usepackage{nicefrac}       
\usepackage{microtype}      
\usepackage{graphicx}
\usepackage{natbib}
\usepackage{doi}
\usepackage{pdfpages}
\usepackage{amssymb, bm}
\usepackage{amsmath}
\usepackage{enumerate}
\usepackage{amsthm}
\usepackage{moreverb}
\usepackage{multirow}
\usepackage{subcaption}
\usepackage{float}
\usepackage{relsize}
\usepackage{geometry}
\usepackage{adjustbox}
\usepackage{diagbox, makecell}
\usepackage{rotating}
\usepackage{caption}
\usepackage{quoting}
\usepackage{pdflscape}
\usepackage{afterpage}

\newcommand{\GG}[1]{}
\newcommand{\indep}{\rotatebox[origin=c]{90}{$\models$}}
\newcommand{\edr}{e.d.r }
\newcommand{\X}{\bm{X}}
\newcommand{\e}{\varepsilon}

\newcommand{\be}{\bm{\beta}}

\newcommand{\bet}{\bm{\beta}^\top}
\newcommand{\bX}{\be^{\top} \X}
\newcommand{\bxonek}{\bet_1 \X, ..., \bet_K \X}
\newcommand{\bxone}{\bet_1 \X}
\newcommand{\bxtwo}{\bet_2 \X}

\newcommand{\B}{\boldsymbol{\mathcal{B}}}

\newcommand{\g}{\bm{\gamma}}
\newcommand{\Sb}{\mathcal{S}}
\newcommand{\Si}{\bm{\Sigma}}

\newcommand*\xbar[1]{%
   \hbox{%
     \vbox{%
       \hrule height 0.5pt 
       \kern0.5ex
       \hbox{%
         \kern-0.18em
         \ensuremath{#1}%
         \kern-0.1em
       }%
     }%
   }%
} 

\DeclareMathOperator*{\argmax}{arg\,max}
\DeclareMathOperator*{\argmin}{arg\,min}

\newtheorem{condition}{Condition}[]

\theoremstyle{remark}
\newtheorem{remark}{Remark}

\usepackage{etoolbox}
\patchcmd{\quote}{\rightmargin}{\leftmargin 0em \rightmargin}{}{}
\newcounter{modelcounter}
\newenvironment{model}{\begin{quote}%
    \refstepcounter{modelcounter}%
  \textbf{ Model \arabic{modelcounter}}%
}{%
\end{quote}%
}

\title{On choosing optimal response transformations for dimension reduction}

\author{ 
{Marina Masioti} \\
	Department of Mathematics and Statistics\\
	La Trobe University\\
	Melbourne, VIC 3086, Australia\\
	\texttt{mmasioti@students.ltu.edu.au} \\
	\And
{Luke A. Prendergast} \\
    School of Engineering and Mathematical Sciences \\ 
    La Trobe University\\
	Melbourne, VIC 3086, Australia\\
	\texttt{luke.prendergast@latrobe.edu.au} \\
	\AND
	Amanda Shaker\\
	Department of Mathematics and Statistics\\ 
	La Trobe University\\
	Melbourne, VIC 3086, Australia\\
	\texttt{A.Shaker@latrobe.edu.au} \\
}

\renewcommand{\shorttitle}{Optimal dimension reduction transformations}

\hypersetup{
pdftitle={OTDR},
pdfsubject={q-bio.NC, q-bio.QM},
pdfauthor={Marina Masioti, Luke A. Prendergast, Amanda Shaker},
pdfkeywords={First keyword, Second keyword, More},
}

\begin{document}
\maketitle

\begin{abstract}
It has previously been shown that response transformations can be very effective in improving dimension reduction outcomes for a continuous response. The choice of transformation used can make a big difference in the visualization of the response versus the dimension reduced regressors. In this article, we provide an automated approach for choosing parameters of transformation functions to seek optimal results. A criterion based on an  influence measure between dimension reduction spaces is utilized for choosing the optimal parameter value of the transformation. Since influence measures can be time-consuming for large data sets, two efficient criteria are  also provided. Given that a different transformation may be suitable for each direction required to form the subspace, we also employ an iterative approach to choosing optimal parameter values. Several simulation studies and a real data example highlight the effectiveness of the proposed methods. 
\end{abstract}



\section{Introduction}

With advances in technology and decreases in data storage costs,  we continue to collect more and more data. It is therefore becoming increasingly important to adapt existing methods to large data sets to visualise data sets that contain many attribute variables. Dimension reduction methods have proven to be a popular tool for the visualisation and modeling of multivariate data in a lower-dimensional framework. 

In the setting of a random univariate continuous response variable, $Y \in \mathbb{R}$, and a random $p$-dimensional predictor variable $\X = [ X_1, \dots, X_p]^\top \in \mathbb{R}^p$, \cite{Li91} considered the dimension reduction model (DRM) 
\begin{equation}\label{eq: drm} 
    Y = g\big(\bxonek, \e \big)
\end{equation}
where $g$ is the unknown \textit{link function}, $\be_1, \dots, \be_K$ are linearly independent $p$-dimensional column vectors and $\e$ is the error term independent of $\X$. It should be noted that a more general form of the DRM exists if we assume that $Y$ is independent of $\X$ given $\bxonek$ expressed as $Y \indep \X\ |\ \bxonek$ \citep[see, e.g.][]{Cook98}.  However, we find the DRM in \eqref{eq: drm} convenient when discussing transformations of a continuous predictor.  

The set, $\Sb = \text{span} (\be_1, \dots, \be_K)$ is referred to as the effective dimension reduction (\edr) space and elements of the \edr space are \edr directions. For identifiability, we assume that $\Sb$ is the Central Dimension Reduction subspace \citep[CDRS, e.g.][]{Cook98}, defined as the intersection of all dimension reduction subspaces.

In this setting, the aim of dimension reduction methods is to find a basis for $\Sb$. Since $\be_1, \dots, \be_K$ cannot be uniquely identified given that the link function is unknown, any set of directions $\g_1, \dots, \g_K$ such that span$(\g_1, \dots, \g_K)=\Sb$ is sufficient. When $K < p$, dimension reduction is achieved without loss of information when $\X$ is replaced by $\g_1^\top \X, \dots, \g_K^\top \X$. 

Many dimension reduction methods exist, however there is no consensus as to which method is best.  Performance depends on the unknown link function $g$, the sample size and many other factors.  For some models, some methods can only find a partial basis for $\Sb$ or return very poor estimates of the basis, often due to the form of $g$.  In some such cases transformations of the response have proven to be effective \citep[e.g.][]{Li92,garnham2013}.  The fact that transformations can greatly improve dimension reduction outcomes motivates us to consider an automated approach to choosing parameters of transformation functions that produce optimal results. 

Our automated approach is implemented on two existing dimension reduction methods, which will be introduced in Section \ref{sec:DR&IFs}. In Section \ref{sec:OTDR}, we introduce the response transformations and the criteria used to choose the optimal parameter value of the transformation. Simulated comparisons and a real-world example are considered in Section \ref{sec:Simulations} and \ref{sec:RealData}. Finally, concluding remarks are provided in Section \ref{sec:Conclusion} and further research discussed. 
 
\section{Dimension reduction and Influence measures}\label{sec:DR&IFs}

Since the introduction of Sliced Inverse Regression \citep[SIR,][]{Li91}, which is capable of finding a basis for $\Sb$ under some conditions, many other methods have followed. In this article, we focus on Ordinary Least Squares \citep[OLS,][]{BR77, BR83} and Principal Hessian Directions Analysis \citep[PHD,][]{Li92}, both of which are methods that can be used for dimension reduction that can benefit from response transformations. 

Dimension reduction methods seek information regarding the form of the link function $g$ from \eqref{eq: drm} by reducing dimensionality to allow for a visual inspection using a Sufficient Summary Plot \citep[SSP,][]{Cook98}.  The SSP is achieved by plotting $Y$ against the reduced regressors $\g_1^\top \X, \dots, \g_K^\top \X$. In the sample setting, let a sample of $n$ observations be denoted by $\{Y_i, \bm{x}_i\}_{i=1}^n$. Then, an Estimated SSP (ESSP) is a plot of the $Y_i$'s against the $\widehat{\g}_1^\top \bm{x}_i$'s, $\dots$, $\widehat{\g}_K^\top \bm{x}_i$'s, where $\widehat{\g}_1, \dots, \widehat{\g}_K$ are the estimated \edr directions.  So far we have not discussed how we can choose $K$.  This will be done when we discuss PHD shortly.

\subsection{Ordinary Least Squares}

OLS is commonly used in the multiple linear regression setting where $\bm{b} = \Si^{-1} \Si_{xy}$ is the population slope vector, $\Si_{xy}$ is the covariance between $Y$ and $\X$, and $\Si$ the variance-covariance matrix of $\X$. However, \cite{BR77, BR83} showed that OLS can be used as a dimension reduction method when $K = 1$, $\e$ is additive and $\X$ is Gaussian for the model in \eqref{eq: drm}. Then, $\be_1$ can be identified up to a multiplicative scalar, meaning that $\bm{b} = c\be_1$, for some $c\in \mathbb{R}$ as long as $c \neq 0$. 
\cite{li1989} generalised this result without the need for a Gaussian $\X$ or an additive error, only requiring the Linear Design Condition (LDC) to hold:

\begin{condition}[LDC]\label{LDC}
For any $\bm{b}\in \mathbb{R}^p$, $\text{E}\big( \bm{b}^\top \X | \bX \big)$ is linear in $\bX$.
\end{condition}

The LDC holds when $\X$ follows an elliptically symmetric distribution although this is not the only assumption under which it holds. When $p$ is large, \cite{hall1993} showed that the LDC will often approximately hold. Note here that even though OLS can be used for $K>1$, it can only provide one informative direction and therefore a partial basis for $\Sb$. 

Further, \cite{li1989} showed that other linear regression methods with different convex criterion functions can also identify e.d.r. directions. Such methods include robust linear regression methods such as $M$-estimators. 

OLS can perform well with a wide variety of models. However, there are some cases where it can fail. One example is when the underlying relationship between $Y$ and $\X$ is symmetric about the mean of $\bX$, in which case $\Si_{xy} = \text{Cov}(\X, Y) = 0$ and no direction is found. Other examples that result in $\bm{b} = \bm{0}$, or close to $\bm{0}$, are much less apparent. For example, \cite{garnham2013} and \cite{GarnhamAlexandraLouise2014Imdr} provide other examples where OLS fails, and also highlight the benefits of transforming the response. 

\subsection{Principal Hessian Directions}\label{sec:PHD} 

\cite{Li92} used Stein's Lemma \citep[Lemma 4;][]{stein1981} and the Hessian matrix $\mathbf{H}_{\bm{x}}$ to introduce principal Hessian direction (PHD) for identifying e.d.r directions.  Assuming that $\X \sim N_p (\bm{\mu}, \Si)$, where $\bm{\mu}$ and $\Si$ are the mean and variance-covariance matrix of $\X$ respectively, then the average Hessian matrix of $E(Y | \X)$ is given by, 
\begin{equation}\label{hesy}
\xbar{\mathbf{H}}_{\bm{x}} = \Si^{-1}\Si_{yxx}\Si^{-1}
\end{equation}
where $\Si_{yxx} = \text{E} \big[ (Y- \mu_Y)(\X - \bm{\mu})(\X - \bm{\mu})^\top \big]$, where $\mu_Y$ is the mean of $Y$. Then, the eigenvectors that correspond to nonzero eigenvalues of  $\xbar{\mathbf{ H}}_{\bm{x}}$ are elements of $\Sb$.  The PHD estimation process is as follows: 

\begin{description}
\item[Step 1.] Standardise the $\bm{x}_i$'s, so that $\bm{z}_i=\widehat{\Si}^{-1/2}(\bm{x}_i-\overline{\bm{x}})$ $(i=1,\ldots,n)$ where $\overline{\bm{x}}$ and $\widehat{\Si}$ are the sample mean and covariance of the $\bm{x}_i$'s respectively.
\item[Step 2.] Calculate the estimate to the average Hessian matrix on the $\bm{z}$-scale as $$\widehat{\Si}_{yzz}=\frac{1}{n}\sum^n_{i=1}(y_i-\overline{y})\bm{z}_i\bm{z}_i^\top$$ where $\overline{y}$ is the sample mean of the $y_i$'s.
\item[Step 3.] Carry out an eigen-decomposition of $\widehat{\Si}_{y\bm{z}\bm{z}}$ and let $\widehat{\bm{\eta}}_1,\widehat{\bm{\eta}}_2,\ldots,\widehat{\bm{\eta}}_p$ denote the eigenvectors associated with the ordered absolute eigenvalues $|\widehat{\lambda}_1|\geq |\widehat{\lambda}_2| \geq \ldots \geq |\widehat{\lambda}_p| \geq 0.$ 
\item[Step 4.] Return $\widehat{\g}_1=\widehat{\Si}^{-1/2}\widehat{\bm{\eta}}_1,\ldots, \widehat{\g}_K=\widehat{\Si}^{-1/2}\widehat{\bm{\eta}}_K$ as the estimated basis for $\Sb$.
\end{description}

In the above we have assumed that $K$ is known, however this is unlikely to be the case in practice.  Under the normality assumption of $\X$ and assuming the average Hessian matrix is of rank $K$, \cite{Li92} shows that
\begin{equation}
    n\sum^p_{j=K+1}\widehat{\lambda}_j^2\sim 2 \text{Var}(Y)\chi^2_{\text{df}}
\end{equation}
where the degrees of freedom is $\text{df}=(p-K+1)(p-K)/2$.  Hence, this could be used iteratively to decide on a choice of $K$.  This will be discussed more later when we discuss the use of test statistics to guide transformation choices. For other methods on estimating $K$ see, for example, \cite{Cook98a} and \cite{Ferre98}.

PHD is not guaranteed to find all $K$ directions (in which case the rank of the average Hessian matrix is less than $K$) and usually performs well in finding directions that have a non-linear association with the response. E.g, as an extreme example, the average Hessian matrix for the multiple linear regression model is equal to $\mathbf{0}$, so that only by chance can an eigenvector be informative.  \cite{Cook98} referred to this phenomenon as elusive linear trends.

Li also pointed out that adding or subtracting a linear function of the predictor from $Y$ does not change the Hessian matrix, and therefore proposed a variation of PHD where the response is replaced by the OLS residual.  In some cases this residual-based PHD can be more successful, on average, in finding directions it is not expected to find \citep[e.g., the elusive linear trends,][]{Cook98,Smith10}. In this paper we focus on the $y$-based PHD since the transformations we consider help to uncover these directions.  However, the residuals could similarly be transformed and so the work that follows can be directly applied when residuals are used instead.

\cite{Li92} provided an example of where a transformation can greatly benefit the PHD results.  Also, \cite{heng2001study} showed that PHD is adversely affected by large values and that by trimming them improvements can be found.  Hence, transformations that reduce the magnitude of observations relative to others may then also be helpful.

\subsection{Iterative dimension reduction}\label{sec:iterative}

Similarly as noted for PHD above, many dimension reduction methods can find only a partial basis for $\Sb$ for some types of models. \cite{Shaker} introduced an iterative application of dimension reduction methods for estimating the full basis of $\Sb$. They proposed the use of different dimension reduction methods for each iteration to collectively form a basis of $\Sb$. They considered combinations of methods that naturally complement each other, e.g. PHD and OLS (as detailed above), SAVE and SIR (methods based on slicing) etc.   In the iterative approach, the \edr direction obtained by the first method is removed from the dimension reduction matrix derived for the following method. This is done by pre- and post-multiplying the matrix with $\textbf{I}_p - \mathbf{P}$, where $\mathbf{P} = \Si^{1/2} \g_1 \g_1^\top \Si^{1/2}$ is the projection matrix onto $\Si^{1/2} \Sb_1 \subset \Si^{1/2} \Sb$ and $\g_1$ is the first \edr direction recovered \citep[see, Proposition 3.1,][]{Shaker}. This ensures that only new information regarding $\Sb$ is obtained in the second iteration since the second direction will be an element of $\Si^{1/2} \Sb \cap \Si^{1/2} \Sb_1'$, where $\Sb_1'$ is the complement of $\Sb_1$.  

When using OLS first followed by PHD (where we write PHD$|$OLS since PHD is carried out conditional on a direction for OLS having already been found), this equates to carrying out an eigen-decomposition on
\begin{equation}
    (\mathbf{I}-\widehat{\mathbf{b}}_z\widehat{\mathbf{b}}_z^\top)\widehat{\Si}_{yzz}(\mathbf{I}-\widehat{\mathbf{b}}_z\widehat{\mathbf{b}}_z^\top)\label{bz}
\end{equation}
where $\widehat{\mathbf{b}}_z$ is the normalised (i.e. $\|\widehat{\mathbf{b}}_z\|=1$) OLS slope vector for the regression of the $y_i$'s on the $\bm{z}_i$'s. In terms of the OLS slope, $\widehat{\mathbf{b}}$, for the regression of the $y_i$'s on the $\bm{x}_i$'s, $\widehat{\mathbf{b}}_z=\widehat{\Si}^{1/2}\widehat{\mathbf{b}}/\|\widehat{\Si}^{1/2}\widehat{\mathbf{b}}\|$.  Hence, the direction resulting from the decomposition of the above matrix will be orthogonal to that already found by OLS, and a new direction (or directions) can be added to estimate the basis for $\Sb$ by re-standardising with respect to $\widehat{\Si}^{-1/2}$ (as in Step 4 of the PHD algorithm). 


\subsection{Influence measures for dimension reduction}\label{sec:IFs}

The Influence Function \citep[IF;][]{Hampel74} is commonly used for assessing the robustness properties of an estimator and these have been derived and studied in the context of dimension reduction \citep[e.g.][]{PRENDERGAST_2005,Smith10}.  We are specifically interested in influence in the sample setting, where we wish to detect observations whose removal from the sample causes a big change in estimation. The aforementioned theoretical works regarding the IF have lead to the introduction of suitable sample versions for dimension reduction.  

There are some challenges that need to be overcome when thinking of influence in the context of dimension reduction.  We are interested in estimators of \edr directions that collectively estimate the dimension reduction subspace. Two subspaces that are equal in span contain exactly the same information for dimension reduction, and this has lead to several influence functions focusing on spans of directions \citep[e.g., for principal component analysis,][]{Benasseni1990}.  On the other hand, two subspaces can be different in span, yet produce almost identical ESSPs in our setting of dimension reduction.  For example, suppose that the target \edr direction is $\be_1=[1,0,\ldots,0]^\top$ and that we also have another candidate $\bm{\beta}_1^*=[0,1,\ldots,0]^\top$.  These two directions are orthogonal to one another, yet the ESSPs (plots of the $y_i$'s versus the dimension reduced $\bm{x}_i$'s using either of the directions) would be similar if the first two predictor variables were highly correlated.  The amount of information lost if we replaced $\be_1$ by $\be_1^*$ depends on the underlying covariance structure of the predictors.  Hence we need influence measures that can detect changes in the dimension reduced predictor space.

Let $\X_n$ denote the predictor matrix whose $i$th row is $\bm{x}_i^\top$ and let $\widehat{\mathbf b}$ be the estimated OLS slope vector.  Then \cite{Prendergast2008} considered the influence measure
\begin{equation}\label{eq:ri}
    r_i=n^2\left[\frac{1}{\text{cor}^2\left(\X_n\widehat{\mathbf b},\X_n\widehat{\mathbf b}_{(i)}\right)} - 1\right]
\end{equation}
where $\widehat{\mathbf b}_{(i)}$ is the estimated OLS slope without the $i$th observation and cor$^2(\cdot,\cdot)$ denotes the squared correlation between the two arguments.

In the case of $K > 1$, the average squared canonical correlation between the dimension reduced predictors based on estimation with and without the $i$th observation are considered instead. Hence, \cite{Smith10} provided a general form of the influence measure to be used with any $K \geq 1$ and applicable to many dimension reduction methods, including OLS, and specifically used this measure to study robustness of PHD. The relative sample influence version of this measures is,
\begin{equation}\label{eq:rhoi}
    \rho_i = (n-1)^2 \Big[ 1 - r^2_{\widehat{\mathcal{B}}}\big(\widehat{\mathcal{B}}_{(i)}\big)\Big]
\end{equation}
where $r^2_{\widehat{\mathcal{B}}}\big(\widehat{\mathcal{B}}_{(i)}\big)$ is the average of the squared canonical correlations between $\X_n\widehat{\B}$ and $\X_n\widehat{\B}_{(i)} $ (where $\widehat{\B} = [\widehat{\g}_1, \dots, \widehat{\g}_K]$ and $\widehat{\B}_{(i)}$ is the matrix of the $\widehat{\g}_{j(i)}$'s (for $j = 1, \dots, K$) estimated without the $i$th observation). 

Therefore, a large $\rho_i$ is obtained when there is a big difference between the \edr spaces, which means that the $i$th observation has a comparatively large effect on estimation of the \edr space.

\section{Optimal transformations for dimension reduction}\label{sec:OTDR}

We begin this section with a motivating example before introducing the methods and transformations.

\subsection{A motivating example}\label{sec:motivating}

\cite{Li92} used the absolute value transformation to show how simple transformations can improve PHD estimation. In this article, one of the transformations we consider is a one-parameter mean-centered absolute value transformation.  At the model level and for $\mu_Y=E(Y)$, this is defined as $T_1(Y; a) = a(Y - \mu_Y) + (1-a)|Y - \mu_Y|$, where $a = [0,1]$. In the sample setting we can apply this to each $y_i$ $(i=1,\ldots,n)$ where we use the sample mean of the $y_i$'s, $\overline{y}$, as an estimate to $\mu_Y$.

Note that $T_1(Y, 1)$ is just the mean-centered response and so the transformation makes no difference to estimation of the e.d.r space. $T_1(Y, 0)$ is the absolute value of  the mean-centered response and an $a\in (0, 1)$ provides a linear combination of the two. 

PHD is excellent at detecting curvature \cite[see, e.g.][]{Li92, Cook98a} or similarly, non-linearity, and by mean centering the response we can be more hopeful of introducing some additional non-linearity when linear relationships are present.  For example, when all the response values are positive, the absolute value transformation will not change anything. However, by mean centering first thereby shifting the response, then these negative values will be folded back to the positive domain creating non-linearity. 

Consider the model,
\begin{equation}\label{eq:motex}
    Y = 2 + 1.2\times(\bX) + 0.5\e
\end{equation}
where $\X \sim N_{10}(\bm{0}, \bm{I}_{10})$, $\be = [1, 0, -2, 0, \dots, 0]$ is the ten-dimensional \edr direction to be estimated and $\e \sim N(0,1)$ independent of $\X$. As discussed in Section \ref{sec:PHD} we expect PHD to fail at finding an informative \edr direction estimate since the Hessian matrix is equal to $\mathbf{0}$. 

We simulated $n = 200$ observations, denoted $\{y_i,\bm{x}_i\}^{n=200}_{i=1}$ from the model in \eqref{eq:motex} and applied PHD on the $T_1(y_i; a)$'s with different values of $a$ set to $a=0, 0.1, 0.2, \dots, 1$.  This produced 11 estimates of the \edr direction. 

\begin{table}[ht]
\centering
\begin{adjustbox}{width = 1\textwidth}
\begin{tabular}{c|rrrrrrrrrrr}
  \toprule
$a$ & $0$ & 0.1 & 0.2 & 0.3 & 0.4 & 0.5 & 0.6 & 0.7 & 0.8 & 0.9 & 1 \\ 
\midrule
  $\overline{\rho}$ &  $\bm{28.467}$ &  29.264 &  31.468 &  36.265 &  45.725 &  62.898 &  91.630 & 137.438 & 192.406 & 244.913 & 291.968 \\ 
  Cor$^2$ & $\bm{0.898}$ & 0.884 & 0.862 & 0.830 & 0.778 & 0.697 & 0.578 & 0.436 & 0.309 & 0.219 & 0.160 \\ 
   \bottomrule
\end{tabular}
\end{adjustbox}
\caption{Results for each choice of $a$, for $n=200$ observations simulated from the model in \eqref{eq:motex}.  Shown are the average of the sample influence measure, $\overline{\rho}$, for each estimated e.d.r direction, and the squared correlations between the dimension reduced predictors using the true and the estimated direction.}
\end{table}\label{tab: MotExTable}

\begin{figure}[h]
    \centering
    \includegraphics[width = \textwidth, page = 1]{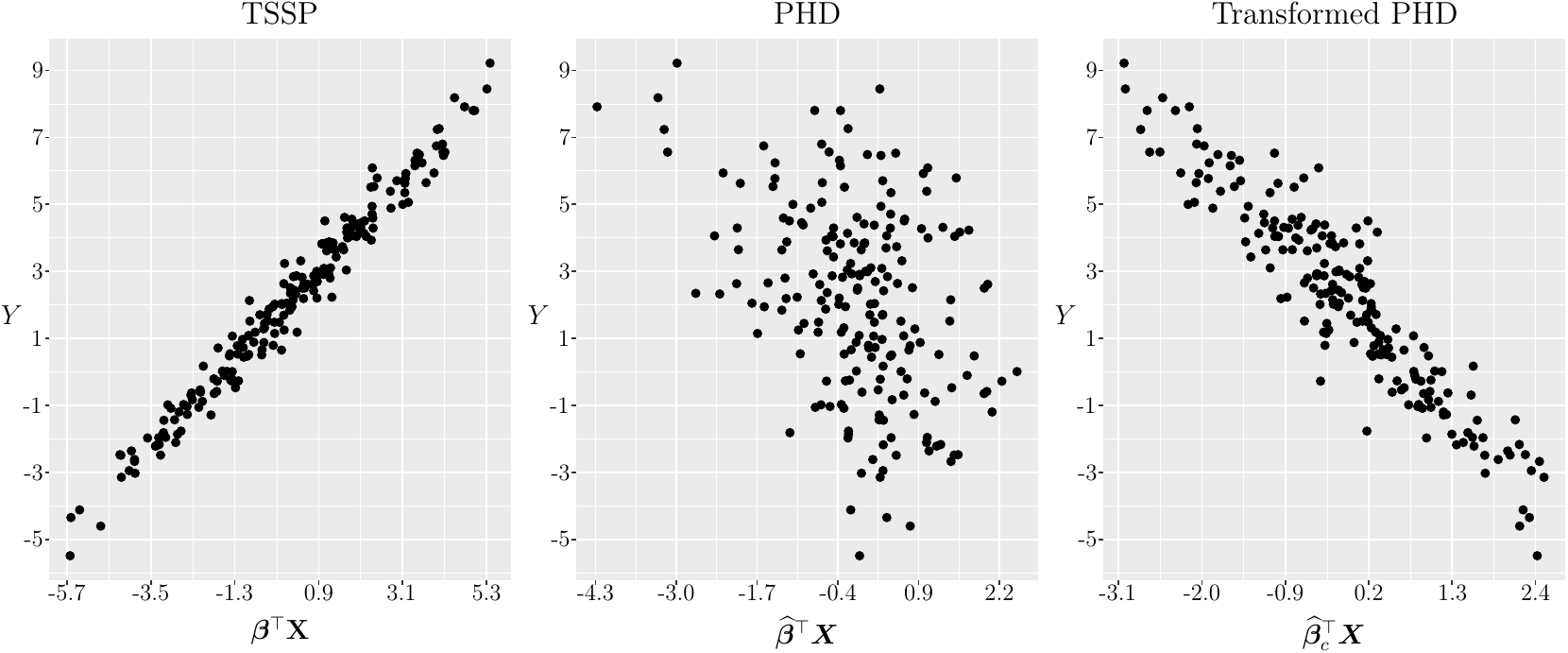}
    \caption{Plots of the true SSP (TSPP; left) of the model in \eqref{eq:motex}, the ESSP using the PHD estimate with no transformation and the ESSP using the PHD estimate (right) with optimal parameter value $a = 0$ for the mean-centered absolute value transformation.}
    \label{fig:MotExESSPs}
\end{figure}

The squared correlation between each $\X_n\widehat{\beta}_a$ and $\X_n\bm{\beta}$ are shown in Table \ref{tab: MotExTable}. Further, the average of the sample influence measure given in \eqref{eq:rhoi} associated with each estimate is shown. The optimal parameter value is the one that results in the largest squared correlation, in this case  choice $a=0$, which results in the absolute value of the mean-centered response.  However, we can only calculate this because, unlike in practice, we know the true e.d.r direction for comparison.  

The choice of $a$ resulting in the largest correlation also resulted in the lowest mean influence, $\overline{\rho}$.  Unlike the correlation, we can compute this sample influence in practice and therefore use this to choose the best $a$.  On average, the estimate of the e.d.r. direction, when $a=0$ is chosen, is the least sensitive to the removal of an observation. The squared correlation decreases as $a$ increases towards one and conversely, the mean sample influence increases. The worst estimate occurs when $a=1$ which still corresponds to a linear model.

Compared to the true SSP (TSSP; left plot in Figure \ref{fig:MotExESSPs}), the ESSP using PHD (middle plot) with no transformation shows that PHD failed to find an informative \edr direction for the model in \eqref{eq:motex}.  However, the PHD estimate using the $T_1(y_i; 0)$'s provides a very good ESSP (right plot) that depicts the linear relationship between $Y$ and $\bX$. Note here, that even though the ESSP shows a mirrored reflection of the TSSP, this simply means that PHD is estimating an e.d.r direction in the direction of $-\be$. 

This motivating example shows that the application of a one-parameter response transformation, along with a criterion for choosing the optimal parameter value of the transformation, can provide great improvements in the estimation of the \edr direction. 

\subsection{A method for optimal parameter selection}

As shown by the motivating example, using a variation of the absolute transformation considered by \cite{Li92}, can greatly improve the estimation of the PHD method. Other works have also further emphasised the benefits of response transformations; for an example of transformations for PHD refer to \cite{Li92}, and to \cite{garnham2013} for a discussion of transformations for OLS. \cite{GarnhamAlexandraLouise2014Imdr} provides an extended discussion of transformations for both PHD and OLS. 

\subsection*{Finding a single direction}

Similar to the motivating example of Section \ref{sec:motivating} using an absolute value transformation and a criterion of minimal mean influence, we consider a general framework for any transformation and criteria.

Let $T(\cdot ; c)$ denote a transformation function with parameter value $c\in \mathbb{R}$. For the single index model with $K = 1$, the algorithm for selecting the optimal parameter value $c$ is straightforward:
\begin{description}
\item[Step 1:] Transform the $y_i$'s using $T(y_i ; c)$ for chosen values of $c$. 
\item[Step 2:] Perform dimension reduction on the transformed responses and $\X_n$, and obtain $\widehat{\be}_{c}$ for each $c$ in Step 1.
\item[Step 3:] Select the optimal $c$ based on a chosen criterion (e.g. based on minimal mean influence) and denote that $c$ as $c^*$. 
\item[Step 4:] Return $\widehat{\be}_1 = \widehat{\be}_{c^*}$ as the estimated e.d.r. direction. 
\end{description}

\subsection*{Finding multiple directions}

The above method can also be used for $K > 1$, where instead of returning only the first \edr direction, we return the first two or more estimates. However, our initial explorations highlighted that improvements could be obtained using $K$ transformations to find $K$ directions, instead of a single transformation of the response (i.e. a single choice of the parameter $c$).   Additionally, a single transformation may only find a partial basis for $\Sb$. Therefore, for $K > 1$, we adopt the iterative approach from \cite{Shaker} to perform iterations of the above method where either the same or different dimension reduction methods can be used for each iteration. Furthermore, the same or different response transformations can be implemented in each iteration. 

Below we assume that PHD is the method to be used to obtain a second direction, although suitable variations of this are possible.
As an example, the iterative approach with $K = 2$ algorithm continues as follows: 
\begin{description}
\item[Step 5:] Calculate the $T(Y; c)$ for chosen values of $c$, where $T$ (and the choices of $c$) can be the same or different to that used for the first iteration. 
\item[Step 6:] Let $\widehat{\be}_1$ be the \edr direction estimated in the first iteration. Then, 
    \begin{description}
    \item[Step 6.1:] Calculate $\widehat{\Si}_{t(y;c)zz}^*=\mathbf{Q} \widehat{\Si}_{t(y;c)zz} \mathbf{Q}$, where $\mathbf{Q}$ is a projection matrix used to remove the component already obtained (see Section \ref{sec:iterative} and Remark \ref{rem:P} below). 
    \item[Step 6.2:] Obtain $\widehat{\be}_{c}=\Si^{-1/2}\widehat{\bm{\eta}}_{c,1}$ for each value of $c$ where $\widehat{\bm{\eta}}_{c,1}$ is the eigenvector that corresponds to the largest non-zero eigenvalue of $\widehat{\Si}_{t(y;c)zz}^*$.
    \end{description}
    \item[Step 7:] Using a chosen criteria, determine the optimal transformation parameter value denoted $c'$. 
    \item[Step 8:] Return $\widehat{\be}_{2}=\widehat{\be}_{c',1}$ as the estimated second e.d.r direction. 
\end{description}

\begin{remark}\label{rem:P}
For $\mathbf{Q} = \mathbf{I}_p - \mathbf{P}$ to be a projection matrix, we need to ensure that it is idempotent, or in this case, ensure that $\|\widehat{\Si}^{1/2} \widehat{\be}_1\|=1$.  If OLS is used first, then $\mathbf{P} = \widehat{\mathbf{b}}_z \widehat{\mathbf{b}}_z^\top$ (see Eqn. \ref{bz}).  If PHD is used first, then $\mathbf{P} = \widehat{\bm{\eta}}_1\widehat{\bm{\eta}}_1^\top$ where $\widehat{\bm{\eta}}_1$ is the eigenvector corresponding to the largest eigenvalue in Step 3 of the PHD algorithm (where, in our approach, $\widehat{\mathbf{b}}_z$ and  $\widehat{\bm{\eta}}_1$ were produced using the appropriate optimally transformed responses). 
\end{remark}

Note that this algorithm searches for each direction in turn. A more exhaustive search can also be implemented: for example, for each possible pair of ${c,c'}$ where these are values for the transformation parameters, compute every pair of estimated e.d.r. directions using the iterative dimension reduction approach and choosing the pair based on a criterion.  However, this is very time-consuming, especially in the high-dimensional setting and our explorations did not reveal any substantial improvements.  

Finally, we limited the approaches above to $K=1$ or $K=2$.  These choices of $K$ allow for visualisation of the response versus the dimension reduced predictors using scatterplots.  However, it is possible that $K >2$, in which case Steps 5 to 7 can be repeated but where $\mathbf{Q}$ is used to remove all previous components.  More on this can be found in \cite{Shaker}.

\subsection{Response Transformations}

Response transformations do not affect Condition \ref{LDC}, required for OLS, or the normality condition of the predictors for PHD. Hence, the following transformations, and others, can be used for the improvement of the \edr direction estimates of the OLS and PHD methods.  When defining the transformations we do so with respect to the random response $Y$ and note that in practice these are applied to the observed responses, $y_i$'s, and where appropriate the sample mean of the $y_i$'s, $\overline{y}$, is used as an estimate to $\mu_y=E(Y)$.  

\subsubsection*{Box-Cox transformation} 

OLS was conceptualized in the setting of linear models, and it was not until later that it was realised it could also be used in dimension reduction for many more models \citep[e.g.][]{li1989}.  However, seeking to linearize the response has been shown to benefit OLS estimation in many cases, and we can be hopeful that such transformations are of benefit in the dimension reduction framework.  To this end we consider the Box-Cox transformation \citep{box1964analysis} given as:
\begin{equation}
    \text{BC}\big(Y; \omega \big) = \begin{cases}
    \text{log}(Y), & \quad \hspace{0.7cm} \omega = 0 \vspace{0.2cm}\\
    \displaystyle\frac{Y^\omega - 1}{\omega},& \quad \hspace{0.7cm} \omega \neq 0
    \end{cases}, \hspace{2cm} \text{where } \omega = [-2, 2].
\end{equation}
Note that we have limited the parameter $\omega\in [-2, 2]$ since we found that the best choice was usually in this range.  It should also be pointed out that this transformation does not help to improve performance when the problem of symmetric dependency occurs.  For example, if $Y$ is symmetric about the mean of $\bm{\beta}^\top\X$, such as when $Y=(\bm{\beta}^\top\X)^2$ where $\X\sim N(\mathbf{0},\Si)$, then the OLS vector is equal to $\mathbf{0}$.  The BC transformation will not fix this problem, although other transformations are possible in this situation \citep{Prendergast2016ResponseAP}.

\subsubsection*{Mean-centered absolute transformation}
For convenience we restate the transformation used in our motivating example  (Section \ref{sec:motivating}):

\begin{equation}
    T_1(Y; c) = c(Y-\mu_Y) + (1-c)\mid Y - \mu_Y\mid, \hspace{1.2cm} \text{where } c = [0, 1].
\end{equation} 

\subsubsection*{Mean-centered absolute Box-Cox transformation}
Recalling that PHD does not like linear trends, we combine elements of the above two transformations to first introduce some possible element of linearity, before applying the absolute mean-centered transformation to benefit PHD. This is given as: 
\begin{equation}
    T_2\big(Y; \omega \big) =  \begin{cases}
    \vspace{0.3cm} \Bigl\lvert\log(Y) - \text{E}\big[\log(Y)\big] \Bigr\rvert, \hspace{1.3cm} \text{$\omega = 0$}  \vspace{0.3cm} \\ 
    
\left|\dfrac{Y^{\omega} - 1}{\omega} - \text{E}\Bigg(\dfrac{Y^{\omega} - 1}{\omega}\Bigg)\right|, \hspace{0.8cm} \text{$\omega \neq 0$} \vspace{0.1cm}
\end{cases}, \quad \hspace{0.3cm} \text{where } \omega = [-2, 2]. 
\end{equation}


For convenience in what follows, we use the following to identify methods using the above transformations.

\begin{description}
\item[BC-OLS:] OLS using the Box-Cox transformation.
\item[$T_1$-PHD:] PHD using the mean-centered absolute transformation ($T_1$).
\item[$T_2$-PHD:] PHD using the mean-centered Box-Cox transformation $(T_2)$.
\item[$T_k$-PHD$|$BC-OLS:] PHD for the second e.d.r. direction using transformation $T_k$ $(k=1,2)$, conditional on the first direction found by BC-OLS.
\item[$T_k$-PHD$|$ $T_j$-PHD:] PHD for the second e.d.r. direction using transformation $T_k$ $(k=1,2)$, conditional on the first direction found by PHD using transformation $T_j$ $(j=1,2)$.
\end{description}

\subsection{Criteria for choosing the optimal parameter value}\label{criteriasec}

We now introduce three criteria that can be used to choose the optimal parameter value of a transformation. 

\subsubsection*{Minimum influence criterion}

In Section \ref{sec:IFs} we presented the generalised sample influence measures derived by \cite{Smith10} for dimension reduction with many methods including OLS and PHD and for any $K$. Then in Section \ref{sec:motivating} we chose the optimal parameter for the mean-centered absolute transformation using the mean influence across all observations. The first criterion uses this influence measure to determine the optimal parameter value of a transformation.

Let us consider BC-OLS. In this case, we let $\bm{\omega}=\{\omega_1, \omega_2,\ldots,\}$ denote the set of transformation parameter values to be used and Step 3 of the algorithm for the single-index model ($K = 1$) becomes:
\begin{description}
\item[Step 3:] Let $\overline{\rho}_\omega$ denote the average of the $\rho_i$ influence values in \eqref{eq:rhoi} where the OLS slope vectors are estimated using the BC transformed $y_i$'s with parameter value $\omega$. Then the optimal parameter value is 
$$\omega' = \argmin_{\omega \in \bm{\omega}} \overline{\rho}_\omega.$$ 
\end{description}
Step 3 is the same for the $T_1$-PHD and $T_2$-PHD methods. 

This also extends to the iterative methods (for $K>1$) where
Step 7 is performed in the same way.  Note that we only have to consider influence associated with the new direction found at this step. This is because the first direction is fixed, so that the first squared canonical correlation from (\ref{eq:rhoi}) will be trivially equal to one since the first direction appears identically in both $\X_n\widehat{\B}$ and $\X_n\widehat{\B}_{(i)} $. We also provide the following comment on the use of the influence measures. 

A disadvantage of using the minimum influence criterion is that many leave-one-out computations are required.  The influence can be computed quicker for OLS even for large data sets compared to the  PHD which involves an eigen-decomposition each time an observation is removed.  If we are doing this for $d$ choices of the transformation parameter, then $(n+1)\times d$ total eigen-decompositions are required (for each direction to be found).  \cite{Smith10} provide an empirical influence measure that quickly approximates $\rho_i$ in \eqref{eq:rhoi}.  However, the approximation is poor when the method performs poorly and so is unreliable in this setting where some transformations may result in poor estimates.  This means that the empirical influence measure cannot be used for its efficiency and that the $\rho_i$'s should be used. Therefore, we also introduce two time-efficient criteria that do not require leave-one-out computations.  

Furthermore, \cite{GarnhamAlexandraLouise2014Imdr} derived the theoretical influence diagnostic for the OLS \edr space following a response transformation and provided an efficient empirical version that can be used in practice to approximate the OLS $\rho_i$'s under the response transformation setting. This measure can also be used in our approach to chose the optimal parameter value of the transformation. 

\subsubsection*{Maximum eigenvalue ratio criterion}

We take advantage of the nature of the eigenvalues of the average Hessian matrix to provide a time-efficient criterion for choosing the optimal parameter value of a transformation. We determine the optimal parameter value by identifying the maximum eigenvalue ratio. Consider, for example, the $T_1(Y; c)$ transformation and let $|\widehat{\lambda}_{c,1}|\geq \ldots |\widehat{\lambda}_{c,p}|\geq 0$ denote the PHD eigenvalues based on estimation with the $T_1(y_i;c)$'s. Then, the ratio of the sum of the $K$ eigenvalues associated with the e.d.r. direction estimators, to the sum of all the eigenvalues, for a given parameter value $c$ is, 
$$\Lambda_{c} = \frac{\sum_{i = 1}^K |\widehat{\lambda}_{c,i}|}{\sum^{p}_{j = 1}|\widehat{\lambda}_{c,j}|}$$
where the $\lambda_c$'s are the eigenvalues estimated from the PHD using the $T_1(y_i; c)$'s with a particular value of $c$.  Larger ratios indicate the e.d.r. directions that are more prominent in the eigen-decomposition of the Hessian matrix. Then the optimal parameter value is, 
$$c' = \argmax_{c \in [0,1]} \Lambda_c.$$

While this approach could be used to simultaneously estimate more than one e.d.r. direction (e.g. choose $K=2$ to find a single transformation that results in the largest two eigenvalues relative to the sum of all eigenvalues), recall that we found different transformations were needed to improve estimation of individual directions.  Hence, in our algorithms we choose $K=1$ either to find a single direction, or to iteratively seek more than one direction.

\subsubsection*{Maximum evidence criterion}

Another efficient way to seek the optimal parameter value is to choose the parameter that maximizes a test statistic used as evidence against $K=0$.   First, we consider a standard test for determining $K$ when using PHD.  Tested sequentially over $k = 0,\ldots, p$, the hypotheses are defined to be
$$H_0:K \leq k\;\;\text{versus}\;\;H_1:K > k,$$
where the number of e.d.r. directions to form the basis is the first $k$ for which $H_0$ is not rejected.  From Theorem 4.2 of \cite{Li92}, a test statistic for a fixed $k$ where $s^2_y$ denotes the sample variance of the $y_i$'s is
\begin{equation}\label{test_stat}
    t_k = \frac{n}{2s^2_y}\sum^p_{j=k+1}\widehat{\lambda}_j^2 \sim \chi^2_{(p-k+1)(p-k)/2}\;\;\text{under}\;\;H_0.
\end{equation}

Hence, to start and for the $T_2$ transformation as an example, let $t^\omega_0$ (i.e. $k=0$ so that $H_0:K = 0$) denote the test statistic above but where PHD estimation has been carried out with the $T_2(y_i; \omega)$'s.  Then
$$\omega' = \argmax_\omega t^\omega_0$$
so that we choose the transformation that maximises the evidence in favour of there being at least one e.d.r. direction.  The most prominent PHD direction can then be used either as the only direction for a $K=1$ model or as the next direction in the iterative dimension reduction.

\section{Simulations}\label{sec:Simulations}

In this section we present simulated examples that demonstrate the performance of the proposed methods in estimating the \edr direction(s) of the OLS and PHD methods. 

For assessing the performance of the dimension reduction methods used in this section, we provide tables of the average squared correlations, in the case of $K=1$, and average squared canonical correlations, in the case of $K=2$, between the true and estimated dimension reduced regressors. In addition, we provide the boxplots of the squared correlations or the squared canonical correlations, for $K= 1$ and $K=2$ respectively, for all of the methods compared in each example. For each model we perform 1000 simulated runs for each combination of the different $n$ and $p$ values chosen. For all examples, we simulate $\X \sim N_p(\bm{0}, \bm{I}_p)$ and $\e \sim N(0,1)$ independent of $\X$.

\subsection{Single-index models}

For dimension reduction with $K = 1$, we consider the following models, 
\begin{model}\label{Model1} 
\quad $Y = 2 \exp{(1 + 1.2 \bX + 0.5\e)^{1/2}} + 0.3\e$ with $\be = [1, 0, 1.5, 0, 0.5, 0, \dots, 0]^\top$.
\end{model}

\begin{model}\label{Model2}
\quad $Y = 1.5\sin(0.7 \bX + 0.25\e) $ with $\be = [1, 0 ,-1, 0.5, 0, \dots, 0]^\top$.
\end{model}

For Model \ref{Model1}, we perform $1000$ simulated runs for each combination of $n = 50, 200, 500$ and $1000$ and $p = 5, 10,$ and $20$. We compared the performance of all the methods for $K = 1$ focusing on the OLS and BC-OLS methods. For each run of Model \ref{Model1}, the BC-OLS method was performed with parameter values from $\omega = -2$ to $\omega = 2$ in increments of $0.1$ and the minimum influence criterion was used to choose the optimal parameter value. 

\begin{table}[h]
\centering
\addtolength{\tabcolsep}{7pt}
\begin{tabular}{ll|rr}
  \toprule
n & p & \makecell{OLS} & \makecell{BC-OLS} \\ 
  \toprule
50 & 5 & 0.864 ( 0.093 ) & 0.987 ( 0.010 ) \\ 
 & 10 & 0.734 ( 0.113 ) & 0.971 ( 0.016 ) \\ 
 & 20 & 0.569 ( 0.116 ) & 0.939 ( 0.026 ) \\ 
 \midrule
200 & 5 & 0.926 ( 0.060 ) & 0.997 ( 0.002 ) \\ 
 & 10 & 0.848 ( 0.086 ) & 0.993 ( 0.003 ) \\ 
 & 20 & 0.732 ( 0.108 ) & 0.986 ( 0.005 ) \\ 
 \midrule
500 & 5 & 0.953 ( 0.044 ) & 0.999 ( 0.001 ) \\ 
 & 10 & 0.904 ( 0.062 ) & 0.997 ( 0.001 ) \\ 
 & 20 & 0.819 ( 0.091 ) & 0.994 ( 0.002 ) \\
 \midrule
1000 & 5 & 0.969 ( 0.032 ) & 0.999 ( 0.000 ) \\ 
 & 10 & 0.935 ( 0.051 ) & 0.999 ( 0.001 ) \\ 
 & 20 & 0.869 ( 0.078 ) & 0.997 ( 0.001 ) \\ 
   \bottomrule
\end{tabular}
\caption{Table of the average squared correlations, cor$^2(\X_n\bm{\beta}, \X_n\widehat{\mathbf{b}})$, for the OLS and BC-OLS methods, across 1000 simulated runs from Model \ref{Model1} for the different choices of $n$ (50, 200, 500 and 1000) and $p$ (5, 10 and 20). The associated standard deviations are shown in parentheses. }
\label{Mod1OLStab}
\end{table}

For OLS, we observe that the average squared correlations in Table \ref{Mod1OLStab} decrease when dimensionality increases. However, as sample size increases the performance of OLS also improves, on average. This indicates that OLS performance declines under the high-dimensional, low sample-size setting and generally can be more sensitive when $p$ is large. 

For BC-OLS, the average squared correlations in Table \ref{Mod1OLStab} show that the method performs extremely well for this model across all choices of $n$ and $p$. Furthermore, changes in the average squared correlations, as dimensionality increases, are much smaller than the OLS method, indicating that BC-OLS is less sensitive to large values of $p$. The standard deviations also show that there is very small estimator variability in the results of BC-OLS which highlights consistently good estimates of the direction. 

\begin{figure}[h!]
    \centering
    \includegraphics[width = 0.8\textwidth, height = 0.85\textheight, page = 1 ,keepaspectratio]{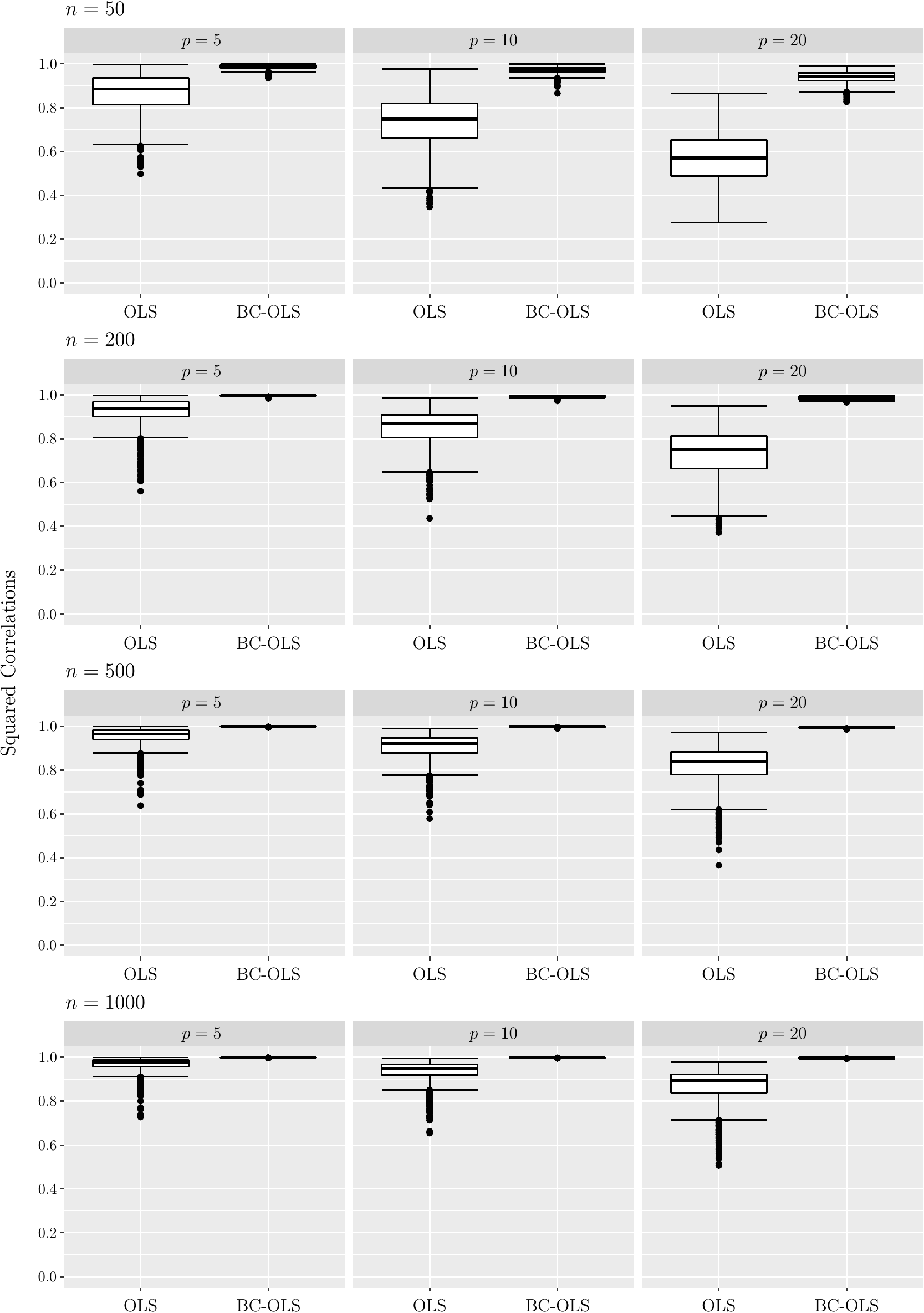}
    \caption{Boxplots of the squared correlations, cor$^2(\X_n\bm{\beta}, \X_n\widehat{\mathbf{b}})$, between the true and the estimated dimension reduced regressors given by the OLS and BC-OLS methods, across 1000 simulated runs from Model \ref{Model1}, for each combination of $n = 50, 200, 500$ and 1000 and $p = 5, 10$ and 20. }
    \label{fig:ExpK1OLSbox}
\end{figure}

The boxplots of the squared correlations for the OLS and BC-OLS methods in Figure \ref{fig:ExpK1OLSbox} support the findings of Table \ref{Mod1OLStab}. Even though OLS can perform well, especially for small $p$, the BC-OLS method provides obvious improvements for Model \ref{Model1}, for each combination of the $n$ and $p$ choices. Furthermore, the greater variability of the OLS correlations is clearly shown indicating that consistently good estimates are not achieved when compared to the BC-OLS method. 

In Table \ref{Mod1-OptValstab} we report the frequency by which a specific value of $\omega$ was chosen as optimal, by the minimum influence criterion for the BC-OLS method, across the 1000 simulated runs from Model \ref{Model1} and for each combination of $n$ and $p$ values. There is decreased variability of the optimal parameter values chosen for $n = 50$ versus the other choices of $n$ and also a shift towards a value between 0 to $-0.2$ as $n$ increases.  Hence, something very close to the log transformation is typically chosen as optimal. 


\begin{table}[h]
\centering
\begin{adjustbox}{width=0.92\textwidth, height = 0.3\textheight, keepaspectratio}
\addtolength{\tabcolsep}{5pt}
\begin{tabular}{r|rrr|rrr|rrr|rrr}
  \toprule
  \makecell{$n$} & & \makecell{50} & & & \makecell{200} & & & \makecell{500} & & &  \makecell{1000}\\ 
  \midrule
 \diagbox{$\omega$}{$p$}  & \makecell{5} & \makecell{10} & \makecell{20} & \makecell{5} & \makecell{10} & \makecell{20} & \makecell{5} & \makecell{10} & \makecell{20} & \makecell{5} & \makecell{10} & \makecell{20} \\
  \midrule
  $\leqslant -0.4$ & 0 & 0 & 0 & 0  & 0 & 0 & 0 & 0 & 0 & 0 & 0 & 0 \\   
 $-0.3$ &  0 & 0 & 0 & 0 & 0 & 0 & 1 & 0 & 0 & 9 & 4 & 4 \\ 
  $-0.2$ &  13 & 2 & 1 & 46 & 31 & 15 & 177 & 151 & 127 & 393 & 393 & 371 \\ 
  $-0.1$ &  154 & 112 & 124 & 597 & 590 & 636 & 775 & 801 & 848 & 595 & 602 & 625 \\ 
  0.0 &  531 & 571 & 536 & 357 & 377 & 349 & 47 & 48 & 25 & 3 & 1 & 0 \\ 
  0.1 &  271 & 286 & 305 & 0 & 2 & 0 & 0 & 0 & 0 & 0 & 0 & 0 \\ 
  0.2 &  29 & 29 & 34 & 0 & 0 & 0 & 0 & 0 & 0 & 0 & 0 & 0 \\ 
  0.3 &  2 & 0 & 0 & 0 & 0 & 0 & 0 & 0 & 0 & 0 & 0 & 0 \\ 
  $\geqslant 0.4$ & 0 & 0 & 0 & 0 & 0 & 0 & 0 & 0 & 0 & 0 & 0 & 0 \\
   \bottomrule
\end{tabular}
\end{adjustbox}
\caption{Counts of the chosen Optimal parameter values, $\omega$, for the BC-OLS method, across 1000 simulated runs from Model \ref{Model1}, for the different choices of $n$ and $p$.}
\label{Mod1-OptValstab}
\end{table}


For Model \ref{Model2}, we performed 1000 simulated runs for each combination of the different choices of $n = 100, 500$ and 1000 and $p = 5, 10$ and 20. We compared the results of the default against the proposed methods for $K =1$, focusing on the comparison between the PHD, PHD with transformation $T_1$ ($T_1$-PHD) and with transformation $T_2$ ($T_2$-PHD).   

Both $T_1$-PHD and $T_2$-PHD provided substantial improvements and similar results, on average. Therefore, for simplicity we only report and discuss the results of $T_1$-PHD and note that there was a little more variability when using the $T_2$ transformation. Note that in each run of the $T_1$-PHD method, the transformation is performed with parameter values from $c = 0$ to $c= 1$ in increments of 0.1. Throughout, we will denote the $T_1$-PHD method applied with each of the aforementioned criteria by $T_1$-PHD$_{\rho}$ (minimum influence), $T_1$-PHD$_{\Lambda}$ (maximum eigenvalue ratio) and $T_1$-PHD$_{t_k}$ (maximum evidence), respectively. 

\begin{table}[h]
\centering
\addtolength{\tabcolsep}{4pt}
\begin{tabular}{lr|llll}
  \toprule
\makecell{$n$} & $p$ & \makecell{PHD} & \makecell{$T_1$-PHD$_{\rho}$} & \makecell{$T_1$-PHD$_{\Lambda}$} & \makecell{$T_1$-PHD$_{t_k}$} \\ 
  \midrule
100 & 5 & 0.174 ( 0.215 ) & 0.850 ( 0.250 ) & 0.769 ( 0.317 ) & 0.865 ( 0.208 ) \\ 
 & 10 & 0.070 ( 0.101 ) & 0.649 ( 0.350 ) & 0.575 ( 0.370 ) & 0.637 ( 0.334 ) \\ 
 & 20 & 0.031 ( 0.049 ) & 0.235 ( 0.305 ) & 0.150 ( 0.234 ) & 0.215 ( 0.289 ) \\ 
 \midrule
500 & 5 & 0.167 ( 0.223 ) & 0.985 ( 0.013 ) & 0.979 ( 0.055 ) & 0.984 ( 0.013 ) \\ 
 & 10 & 0.068 ( 0.098 ) & 0.963 ( 0.022 ) & 0.962 ( 0.023 ) & 0.963 ( 0.023 ) \\ 
 & 20 & 0.029 ( 0.044 ) & 0.920 ( 0.050 ) & 0.917 ( 0.059 ) & 0.912 ( 0.078 ) \\ 
 \midrule
1000 & 5 & 0.174 ( 0.222 ) & 0.992 ( 0.006 ) & 0.992 ( 0.007 ) & 0.992 ( 0.006 ) \\ 
 & 10 & 0.069 ( 0.107 ) & 0.983 ( 0.010 ) & 0.982 ( 0.010 ) & 0.983 ( 0.010 ) \\ 
 & 20 & 0.024 ( 0.036 ) & 0.962 ( 0.017 ) & 0.962 ( 0.017 ) & 0.962 ( 0.018 ) \\ 
   \bottomrule
\end{tabular}
\caption{Table of the average squared correlations, cor$^2(\X_n\bm{\beta}, \X_n\widehat{\mathbf{b}})$, for the PHD, $T_1$-PHD$_{\rho}$, $T_1$-PHD$_{\Lambda}$ and $T_1$-PHD$_{t_k}$ methods, for 1000 simulated runs from Model \ref{Model2}, across different choices of $n$ (100, 500 and 1000) and $p$ (5, 10 and 20). Standard deviations are shown in parentheses.}
\label{PHDk1cors}
\end{table}

The average squared correlations in Table \ref{PHDk1cors} show that PHD performs poorly for Model \ref{Model2}, whereas the $T_1$-PHD methods perform extremely well with similar results provided by the different criteria. Generally, PHD does not perform well when the sample-size is small and in higher dimensions. This is true, to a smaller extent, for the $T_1$-PHD methods, which even though they provide good improvements when $n = 100$, they show a greater variability in estimation. For the larger sample sizes of $n=200$ and $n=500$, large improvements are achieved whereas PHD continues to fail.  Furthermore, both the maximum eigenvalue ratio ($\Lambda$) and the maximum evidence ($t_k$) criteria provide results very close to those of the minimum influence criterion for this model. 

\begin{figure}[p]
    \centering
    \includegraphics[width = 0.98\textwidth, page = 1, keepaspectratio]{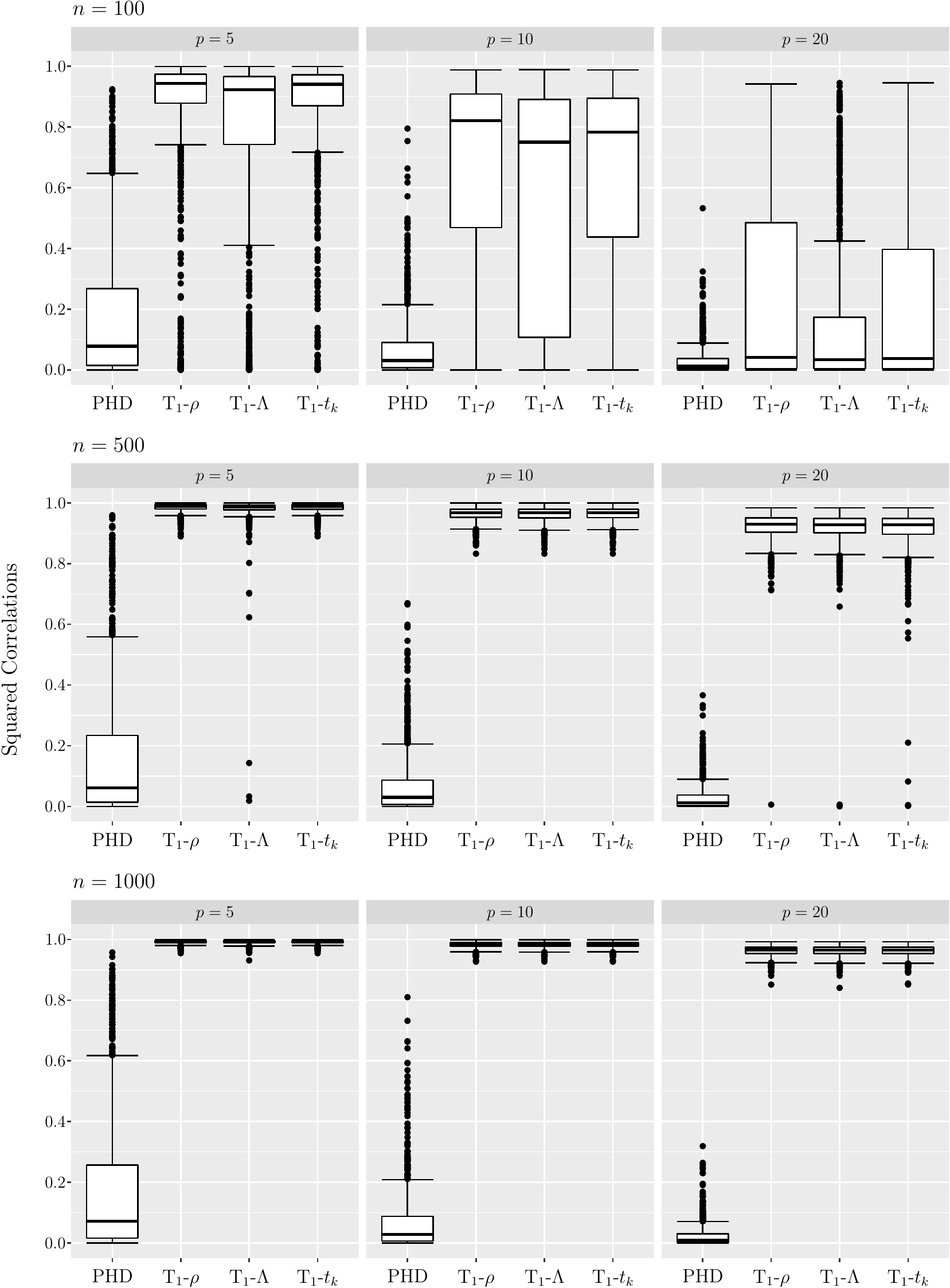}
    \caption{Boxplots of the squared correlations, cor$^2(\X_n\bm{\beta}, \X_n\widehat{\mathbf{b}})$, given by the PHD, $T_1$-PHD$_{\rho}$, $T_1$-PHD$_{\Lambda}$ and $T_1$-PHD$_{t_k}$ methods (denoted by $T_1$-$\rho$, $T_1$-$\Lambda$ and $T_1$-$t_k$ for brevity), across 1000 simulated runs from Model \ref{Model2}, for the different choices of $n$ and $p$.}
    \label{PHDk1Boxplots}
\end{figure}

The boxplots of the PHD, $T_1$-PHD$_{\rho}$, $T_1$-PHD$_{\Lambda}$ and $T_1$-PHD$_{t_k}$ squared correlations in Figure \ref{PHDk1Boxplots} support the above interpretations, clearly showing the variability and reduced performance of the $T_1$-PHD method in the high-dimensional low sample-size setting. They also highlight the big improvements using the $T_1$ response transformation in the estimation of the dimension reduced regressors when compared to PHD. 

Table \ref{OptValsPHDk1} shows the counts of the optimal parameter values chosen from the different criteria ($\rho, \Lambda$ and $t_k$) across 1000 simulated runs from Model \ref{Model2} for the different values of $n$ and $p$ for the $T_1$-PHD method. The counts show greater variability when $n = 100$ which is expected due to PHD's sensitivity to small sample-sizes. However, as $n$ increases the frequencies are concentrated mainly in two choices. The optimal parameter value chosen most often by all three criteria for Model \ref{Model2} is $c = 0$, which means that the optimal transformation will probably be, $T_1(Y; 0) = \mid Y - \mu_y \mid$.

\begin{sidewaystable}
\addtolength{\tabcolsep}{2pt}
\begin{adjustbox}{width = \textwidth, keepaspectratio}
\begin{tabular}{r|rrr|rrr|rrr|rrr|rrr|rrr|rrr|rrr|rrr}
\toprule
$n$ & & \multicolumn{7}{c}{100} & & & \multicolumn{7}{c}{500} & & & \multicolumn{7}{c}{1000} & \\
\midrule
$p$   &  & \makecell{5} & &  & \makecell{$10$} & & &  \makecell{$20$} & &  & \makecell{$5$} & & & \makecell{$10$} & & & \makecell{$20$} & &  & \makecell{5} &  &  & \makecell{10} & & & \makecell{20} &  \\
  \midrule
$c$ & $\rho$ & $\Lambda$ & $t_k$ & $\rho$ & $\Lambda$ & $t_k$ & $\rho$ & $\Lambda$ & $t_k$ & $\rho$ & $\Lambda$ & $t_k$ & $\rho$ & $\Lambda$ & $t_k$ & $\rho$ & $\Lambda$ & $t_k$ & $\rho$ & $\Lambda$ & $t_k$ & $\rho$ & $\Lambda$ & $t_k$ & $\rho$ & $\Lambda$ & $t_k$ \\ 
  \toprule
 0.0 & 631 & 485 & 655 & 555 & 420 & 550 & 289 & 176 & 426 & 814 & 597 & 917 & 866 & 691 & 843 & 945 & 788 & 728 & 914 & 629 & 977 & 950 & 743 & 961 & 960 & 846 & 875 \\ 
 0.1 & 225 & 137 & 220 & 196 & 155 & 198 & 159 & 71 & 142 & 185 & 213 & 83 & 134 & 263 & 155 & 54 & 201 & 226 & 86 & 225 & 23 & 50 & 235 & 39 & 40 & 154 & 124 \\ 
 0.2 & 50 & 116 & 70 & 37 & 110 & 99 & 114 & 80 & 109 & 1 & 119 & 0 & 0 & 45 & 2 & 1 & 8 & 40 & 0 & 109 & 0 & 0 & 22 & 0 & 0 & 0 & 1 \\ 
 0.3 & 18 & 59 & 28 & 28 & 56 & 37 & 52 & 187 & 74 & 0 & 52 & 0 & 0 & 1 & 0 & 0 & 1 & 2 & 0 & 30 & 0 & 0 & 0 & 0 & 0 & 0 & 0 \\ 
0.4 & 9 & 40 & 7 & 31 & 48 & 25 & 50 & 244 & 47 & 0 & 11 & 0 & 0 & 0 & 0 & 0 & 1 & 1 & 0 & 6 & 0 & 0 & 0 & 0 & 0 & 0 & 0 \\ 
 0.5 & 17 & 25 & 3 & 24 & 57 & 12 & 27 & 129 & 17 & 0 & 2 & 0 & 0 & 0 & 0 & 0 & 1 & 3 & 0 & 1 & 0 & 0 & 0 & 0 & 0 & 0 & 0 \\ 
 0.6 & 11 & 26 & 1 & 20 & 30 & 10 & 26 & 31 & 18 & 0 & 1 & 0 & 0 & 0 & 0 & 0 & 0 & 0 & 0 & 0 & 0 & 0 & 0 & 0 & 0 & 0 & 0 \\ 
 0.7 & 6 & 20 & 1 & 19 & 20 & 5 & 38 & 7 & 13 & 0 & 3 & 0 & 0 & 0 & 0 & 0 & 0 & 0 & 0 & 0 & 0 & 0 & 0 & 0 & 0 & 0 & 0 \\ 
 0.8 & 6 & 14 & 15 & 12 & 11 & 1 & 48 & 1 & 6 & 0 & 2 & 0 & 0 & 0 & 0 & 0 & 0 & 0 & 0 & 0 & 0 & 0 & 0 & 0 & 0 & 0 & 0 \\ 
 0.9 & 3 & 18 & 0 & 15 & 8 & 3 & 47 & 1 & 4 & 0 & 0 & 0 & 0 & 0 & 0 & 0 & 0 & 0 & 0 & 0 & 0 & 0 & 0 & 0 & 0 & 0 & 0 \\ 
 1.0 & 24 & 60 & 0 & 63 & 85 & 60 & 150 & 73 & 144 & 0 & 0 & 0 & 0 & 0 & 0 & 0 & 0 & 0 & 0 & 0 & 0 & 0 & 0 & 0 & 0 & 0 & 0 \\ 
   \bottomrule
\end{tabular}
\end{adjustbox}
\caption{Counts of the Optimal parameter values chosen across 1000 simulated runs from Model \ref{Model2}, for the $T_1$-PHD$_{\rho}$, $T_1$-PHD$_{\Lambda}$ and $T_1$-PHD$_{t_k}$ methods for the different combinations of $n = 100, 500$ and $1000$ and $p = 5, 10$ and $20$.}\label{OptValsPHDk1}
\vspace{1cm}
\centering
\begin{tabular}{l|rrr}
  \toprule
$n$ & $T_1$-PHD$_{\rho}$ & $T_1$-PHD$_{\Lambda}$ & $T_1$-PHD$_{t_k}$ \\ 
  \midrule
$100$ & 1.6143 & 0.0250 & 0.0790 \\ 
  $500$ & 26.7699 & 0.0300 & 0.0610 \\ 
  $1000$ & 102.6824 & 0.0200 & 0.1375 \\ 
   \bottomrule
\end{tabular}
\caption{Times taken, in seconds, to perform a single trial of the $T_1$-PHD$_{\rho}$, $T_1$-PHD$_{\Lambda}$ and $T_1$-PHD$_{t_k}$ method with $n = 100, 500$ and $1000$ and $p = 10$, for Model \eqref{Model2}. }\label{tab:times}
\end{sidewaystable}

Finally, Table \ref{tab:times} clearly shows that the time taken to perform the $T_1$-PHD method with the minimum influence criterion ($\rho$) increases rapidly when the sample size increases. Increases in dimensionality would result in further increased times. However, the $T_1$-PHD method with the maximum eigenvalue ($\Lambda$) and maximum evidence ($t_k$) criteria is performed very quickly with very small changes across the different sample sizes. We see that the $\Lambda$ and $t_k$ criteria are very efficient and capable of providing an improved \edr direction estimate. These calculations were performed using an AMD Ryzen 7 2700X Eight-Core Processor 3.70 GHz processor with 32.0 GB RAM using RStudio  Version 1.3.1056 with R version 4.0.2.

\subsection{Multiple-index models}

\cite{Li92} considered the following model to demonstrate how simple response transformations can aid PHD in improving  estimation of the \edr directions. Here, we will use the same model to compare the estimation of the \edr predictors of PHD with and without the proposed iterative response transformations approach. An error term of the appropriate size has been added to the original model to form a more realistic example. So, for dimension reduction with $K=2$, the first model we consider is, 

\begin{model}\label{LisMod}
\quad $Y = \dfrac{1}{3}(\bxone)^3 - (\bxone)(\bxtwo)^2 + 0.4\e$ where $\be_1 = [1, 0, \dots, 0]$, $\be_2 = [0, 1, 0, \dots, 0]$. 
\end{model}

For Model \ref{LisMod} we performed 1000 simulated runs of the PHD, $T_1$-PHD$|$ $T_1$-PHD and $T_2$-PHD$|$ $T_2$-PHD methods, for each combination of $n = 200, 500$ and 1000 and $p = 5, 10$ and 20. We focus on the results given by the default PHD and compare them with the $T_2$-PHD$|$ $T_2$-PHD method which showed the best improvements. For brevity, we will denote the iterative $T_2$-PHD method with each of the criteria, as $T_2$-PHD$^2_{\rho}$, $T_2$-PHD$^2_{\Lambda}$ and $T_2$-PHD$^2_{t_k}$, respectively.

\begin{table}[h!]
\centering
\begin{adjustbox}{width=0.72\textwidth, height = 0.13\textheight, keepaspectratio}
\begin{tabular}{ll|rrrrr}
  \toprule
n & p & \makecell{PHD} & \makecell{$T_2$-PHD$^2_{\rho}$} & \makecell{$T_2$-PHD$^2_{\Lambda}$} & \makecell{$T_2$-PHD$^2_{t_k}$} \\ 
  \midrule
200 & 5 & 0.633 ( 0.325 ) & 0.910 ( 0.143 ) & 0.846 ( 0.179 ) & 0.917 ( 0.115 ) \\ 
  & 10 & 0.490 ( 0.285 ) & 0.805 ( 0.163 ) & 0.648 ( 0.249 ) & 0.763 ( 0.199 ) \\ 
  & 20 & 0.356 ( 0.228 ) & 0.593 ( 0.223 ) & 0.414 ( 0.238 ) & 0.523 ( 0.242 ) \\ 
  \midrule
  500 & 5 & 0.646 ( 0.316 ) & 0.961 ( 0.057 ) & 0.895 ( 0.122 ) & 0.968 ( 0.036 ) \\ 
  & 10 & 0.511 ( 0.295 ) & 0.910 ( 0.079 ) & 0.761 ( 0.200 ) & 0.909 ( 0.078 ) \\ 
  & 20 & 0.398 ( 0.245 ) & 0.813 ( 0.113 ) & 0.527 ( 0.262 ) & 0.755 ( 0.177 ) \\ 
  \midrule
  1000 & 5 & 0.656 ( 0.318 ) & 0.976 ( 0.040 ) & 0.919 ( 0.105 ) & 0.984 ( 0.017 ) \\ 
  & 10 & 0.534 ( 0.283 ) & 0.950 ( 0.051 ) & 0.808 ( 0.181 ) & 0.955 ( 0.037 ) \\ 
  & 20 & 0.429 ( 0.250 ) & 0.892 ( 0.082 ) & 0.621 ( 0.260 ) & 0.884 ( 0.091 ) \\ 
   \bottomrule
\end{tabular}
\end{adjustbox}
\caption{Shows the average of the squared canonical correlations of the PHD, $T_2$-PHD$^2_{\rho}$, $T_2$-PHD$^2_{\Lambda}$ and $T_2$-PHD$^2_{t_k}$ methods (for simplicity the $^2$ superscript indicates that the methods were used twice iteratively to recover the two directions), across 1000 simulated runs from Model \ref{LisMod} for three choices of $n$ (200, 500 and 1000) and three values of $p$ (5, 10 and 20). The corresponding standard deviations are shown in parentheses.}
\label{LisModCorsTab}
\end{table}
\begin{figure}[h!]
    \centering
    \includegraphics[width = \linewidth, height = 0.5\textheight, page = 1, keepaspectratio]{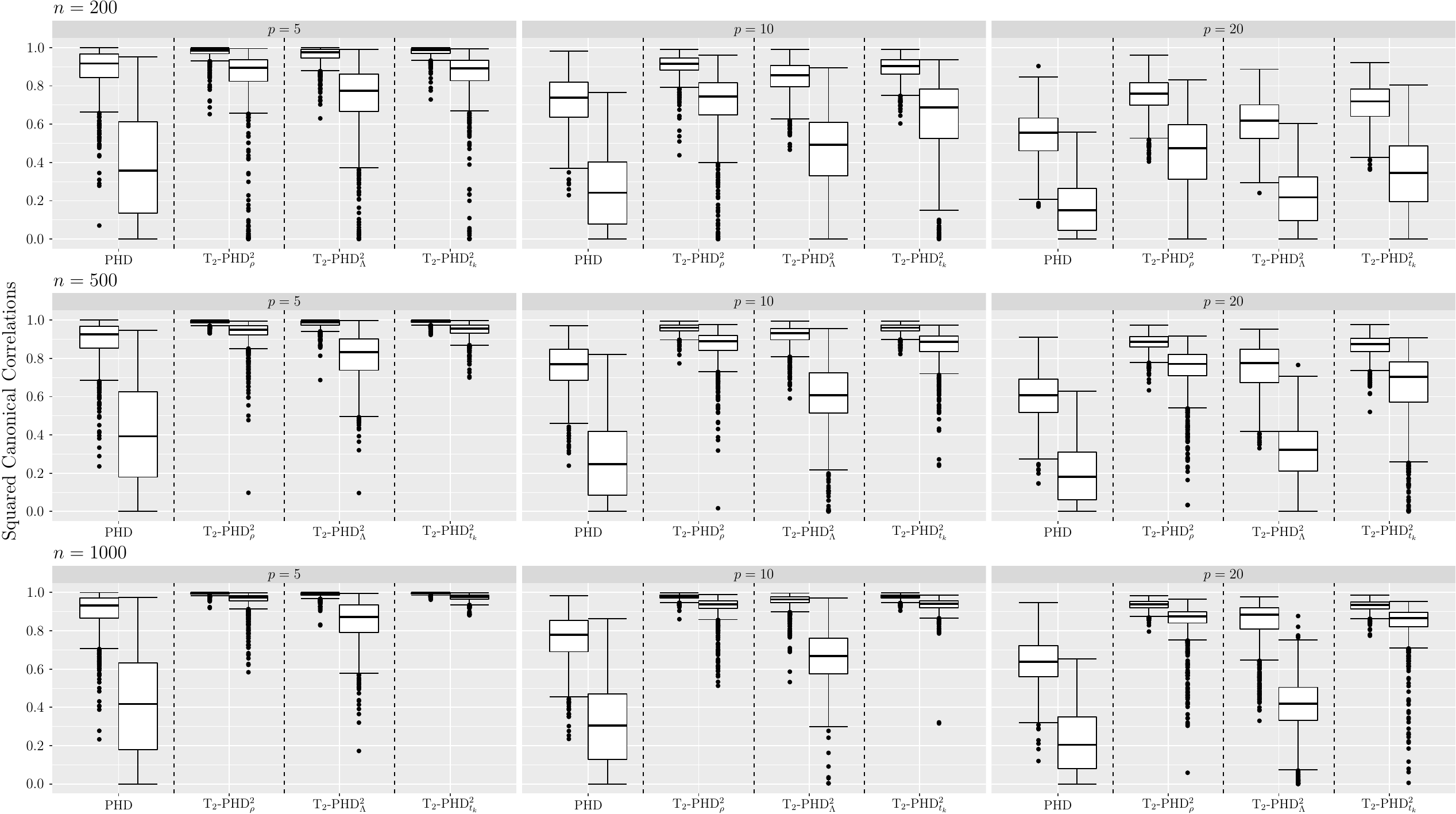}
    \caption{Boxplots of the squared canonical correlations of the PHD, $T_2$-PHD$^2_{\rho}$, $T_2$-PHD$^2_{\Lambda}$ and $T_2$-PHD$^2_{t_k}$ methods  across 1000 simulated runs from Model \ref{LisMod}, for three choices of $n$ (200, 500 and 1000) and three values of $p$ (5, 10 and 20). }
    \label{fig:LisBoxplot}
\end{figure}

Table \ref{LisModCorsTab} shows the average of the squared canonical correlations for each of the aforementioned methods, across 1000 simulated runs from Model \ref{LisMod}, for each combination of the $n$ and $p$ values chosen. The corresponding standard deviations are given in parentheses. The results indicate that the default PHD performs poorly and with greater variability, as seen by the standard deviations, compared to the three variations of the iterative $T_2$-PHD method which provide significant improvements. Furthermore, the maximum evidence ($t_k$) criterion performs slightly better and with less variability than the maximum eigenvalue ratio ($\Lambda$) criterion, and provides similar results to those of the minimum influence ($\rho$) criterion. 

In Figure \ref{fig:LisBoxplot}, we present the boxplots of the squared canonical correlations of each of the methods, across 1000 simulated runs for each choice of $n$ and $p$, to show the performance of the methods in greater detail. The boxplots show that PHD can estimate only a partial basis well, when dimensionality is low, but performance deteriorates for both directions as the dimension increases. 

On the other hand, all three variants of the iterative $T_2$-PHD method show improved estimates for both directions. The results decline as the dimension increases but improve further when sample size increases. As mentioned earlier, the $\Lambda$ and $t_k$ criteria were proposed as alternatives to the $\rho$ criterion for time efficiency and can provide different results for the same model. Here, the $t_k$ criterion is similar to the $\rho$ criterion whereas the $\Lambda$ shows smaller, and in some cases insufficient, improvements in estimation, for Model \ref{LisMod}. 

\begin{table}[h]
\centering
\begin{adjustbox}{width = \textwidth, keepaspectratio}
\begin{tabular}{r|rr|rr|rr||rr|rr|rr||rr|rr|rr}
\toprule
 $p$ & & \multicolumn{4}{c}{5} & & & \multicolumn{4}{c}{10} & & &  \multicolumn{4}{c}{20} & \\
  \midrule
Criterion & \multicolumn{2}{c}{$\rho$} & \multicolumn{2}{c}{$\Lambda$} & \multicolumn{2}{c||}{$t_k$} & \multicolumn{2}{c}{$\rho$} & \multicolumn{2}{c}{$\Lambda$} & \multicolumn{2}{c||}{$t_k$} & \multicolumn{2}{c}{$\rho$} & \multicolumn{2}{c}{$\Lambda$} & \multicolumn{2}{c}{$t_k$} \\ 
  \midrule
  $\omega$ & $\omega_1$ & $\omega_2$ & $\omega_1$ & $\omega_2$ & $\omega_1$ & $\omega_2$ & $\omega_1$ & $\omega_2$ & $\omega_1$ & $\omega_2$ & $\omega_1$ & $\omega_2$ & $\omega_1$ & $\omega_2$ & $\omega_1$ & $\omega_2$ & $\omega_1$ & $\omega_2$ \\
  \toprule
$[-2,-1.6]$ &  72 &   6 & 836 &   2 &   0 &   0 &  34 &   1 & 822 &  16 &   1 &   0 &  26 &   1 & 927 &  77 &  53 &   0 \\ 
  $(-1.6,-1.2]$ &  22 &   2 & 122 &   9 &   0 &   0 &  12 &   0 & 122 &   9 &   2 &   0 &   4 &   0 &  54 &  59 &  10 &   0 \\ 
  $(-1.2,-0.8]$ &  46 &  12 &  12 &  26 &   0 &   0 &  19 &   5 &  11 &  18 &   0 &   0 &   7 &   4 &   1 &  61 &  20 &   0 \\ 
  $(-0.8,-0.4]$ &  56 &  17 &   1 &  74 &   0 &   0 &  30 &   8 &   6 &  38 &   4 &   0 &  16 &   8 &   0 &  55 &  18 &   0 \\ 
  $(-0.4,0]$ &  87 &  60 &   1 &  89 &   0 &   0 &  87 &  19 &   1 &  67 &  11 &   0 &  38 &   8 &   0 &  55 &  42 &   0 \\ 
  $(0,0.4]$ & 107 &  91 &   1 &  88 &   9 &   0 & 106 &  87 &   0 &  83 &  23 &   0 & 112 &  61 &   0 &  38 &  77 &   0 \\ 
  $(0.4,0.8]$ & 108 & 164 &   0 &  93 & 119 &   0 & 175 & 209 &   2 &  69 & 183 &   0 & 207 & 213 &   0 &  24 & 176 &   0 \\ 
  $(0.8,1.2]$ & 110 & 209 &   0 &  73 & 655 &   0 & 144 & 252 &   0 &  70 & 530 &   0 & 169 & 264 &   0 &  14 & 309 &   0 \\ 
  $(1.2,1.6]$ &  75 & 172 &   0 &  89 & 200 &   0 & 125 & 185 &   0 &  50 & 204 &   0 & 167 & 224 &   0 &   7 & 155 &   0 \\ 
  $(1.6,2]$ & 317 & 267 &  27 & 457 &  17 & 1000 & 268 & 234 &  36 & 580 &  42 & 1000 & 254 & 217 &  18 & 610 & 140 & 1000 \\ 
   \bottomrule
\end{tabular}
\end{adjustbox}
\caption{Counts of the optimal parameter values chosen across the 1000 simulated runs from Model \ref{LisMod}, for $n = 500$ and $p = 5, 10$ and 20, from the $T_2$-PHD$^2_{\rho}$, $T_2$-PHD$^2_{\Lambda}$ and $T_2$-PHD$^2_{t_k}$ methods.}
\label{tab:LisOptVals}
\end{table}

In Table \ref{tab:LisOptVals}, we present the counts of the optimal parameter values chosen by the $T_2$-PHD$^2_{\rho}$, $T_2$-PHD$^2_{\Lambda}$ and $T_2$-PHD$^2_{t_k}$ methods, for $n = 500$ and $p = 5, 10$ and 20. In each iteration, the $T_2$-PHD method was performed for $\omega = -2$ to $\omega = 2$ in increments of 0.1. For brevity, we chose to group the parameter value choices in intervals, so that the counts represent the frequency by which parameter values within a particular interval were chosen as optimal across the 1000 simulated runs.   The frequencies highlight the similarities between the $\rho$ and $t_k$ criteria on the choice of optimal parameter values, and the difficulty of the $\Lambda$ criterion in finding the optimal value for one of the \edr directions of Model \ref{LisMod}.  This is also evident from the boxplots of the $T_2$-PHD$^2_{\Lambda}$ method that show a poor performance in finding a good estimate for the second \edr direction. 


The final model we consider is,
\begin{model}\label{SinCubeModk2}
$Y = 5\sin(0.5\bxone) + 0.5 (0.5 \bxtwo)^3 + 0.3\e,$ where $\be_1 = [1, 2, -3, 0, \dots, 0]$, $\be_2 = [1, 1, 0, -2, 0, \dots, 0].$
\end{model}

For Model \ref{SinCubeModk2} we consider the results of the PHD, PHD$|$OLS, $T_1$-PHD$|$ $T_1$-PHD, $T_2$-PHD$|$ $T_2$-PHD, $T_1$-PHD$|$BC-OLS and $T_2$-PHD$|$BC-OLS methods. Each method was performed for 1000 simulated runs from Model \ref{SinCubeModk2}, for each combination of $n = 200, 500$ and 1000, and $p = 5, 10$ and 20. Here, we only report the results of the PHD, PHD$|$OLS and $T_2$-PHD$|$BC-OLS methods where the later provided the biggest improvements compared to the other methods considered for this example. The PHD$|$OLS method is considered to allow for the comparison between the iterative approach \citep{Shaker} and the iterative transformations approach (both of which allow for different dimension reduction methods to be used in each iteration). 
\begin{table}[h]
\centering
\begin{adjustbox}{width=0.95\textwidth, keepaspectratio}
\begin{tabular}{cr|ccccc}
  \toprule
$n$ & $p$ & \makecell{PHD} & \makecell{PHD$|$OLS} & \makecell{$T_2$PHD$_{\rho}$ $|$BC-OLS} & $T_2$PHD$_{\Lambda}$ $|$BC-OLS & $T_2$PHD$_{t_k}$ $|$BC-OLS \\ 
  \midrule
200 & 5 & 0.510 ( 0.379 ) & 0.647 ( 0.393 ) & 0.882 ( 0.181 ) & 0.786 ( 0.284 ) & 0.863 ( 0.202 ) \\ 
 & 10 & 0.345 ( 0.320 ) & 0.561 ( 0.419 ) & 0.781 ( 0.246 ) & 0.658 ( 0.348 ) & 0.750 ( 0.270 ) \\ 
 & 20 & 0.246 ( 0.249 ) & 0.501 ( 0.422 ) & 0.668 ( 0.287 ) & 0.541 ( 0.392 ) & 0.639 ( 0.309 ) \\ 
 \midrule
500 & 5 & 0.512 ( 0.380 ) & 0.657 ( 0.392 ) & 0.939 ( 0.126 ) & 0.842 ( 0.247 ) & 0.925 ( 0.142 ) \\ 
 & 10 & 0.360 ( 0.329 ) & 0.570 ( 0.432 ) & 0.873 ( 0.175 ) & 0.724 ( 0.327 ) & 0.845 ( 0.201 ) \\ 
 & 20 & 0.275 ( 0.276 ) & 0.522 ( 0.448 ) & 0.795 ( 0.218 ) & 0.611 ( 0.385 ) & 0.760 ( 0.250 ) \\
 \midrule
1000 & 5 & 0.523 ( 0.379 ) & 0.654 ( 0.396 ) & 0.963 ( 0.093 ) & 0.871 ( 0.221 ) & 0.950 ( 0.115 ) \\ 
 & 10 & 0.376 ( 0.340 ) & 0.576 ( 0.434 ) & 0.925 ( 0.129 ) & 0.795 ( 0.292 ) & 0.904 ( 0.151 ) \\ 
 & 20 & 0.297 ( 0.292 ) & 0.531 ( 0.456 ) & 0.872 ( 0.157 ) & 0.691 ( 0.359 ) & 0.843 ( 0.187 ) \\ 
   \bottomrule
\end{tabular}
\end{adjustbox}
\caption{Averages of the squared canonical correlations across 1000 simulated runs from Model \ref{SinCubeModk2}, for the PHD, PHD$|$OLS and $T_2$-PHD$|$BC-OLS methods for three choices of $n=$ 200, 500 and 1000 and three values of $p=$ 5, 10 and 20. The corresponding standard deviations are shown in parentheses.}
\label{tab:SinCubeCors}
\end{table}

The results in Table \ref{tab:SinCubeCors} indicate that PHD$|$OLS is an improvement over the default PHD method which performs poorly for Model \ref{SinCubeModk2}. However, PHD$|$OLS still performs poorly, on average, and its performance is consistent across the different sample sizes and dimensions showing high estimator variability. The $T_2$-PHD$_{\rho}$ $|$BC-OLS method has provided clear improvements over the PHD and PHD$|$OLS methods  where performance increases as sample sizes increases. Additionally, the results of the $T_2$-PHD$_{t_k}$ $|$BC-OLS method are similar to those of the $T_2$-PHD$_{\rho}$ $|$BC-OLS and show a better performance compared to the $T_2$-PHD$_{\Lambda}$ $|$BC-OLS with smaller standard deviations. 

\afterpage{
\begin{sidewaysfigure}
    \centering
    \includegraphics[ width = \textwidth, , keepaspectratio, page = 1]{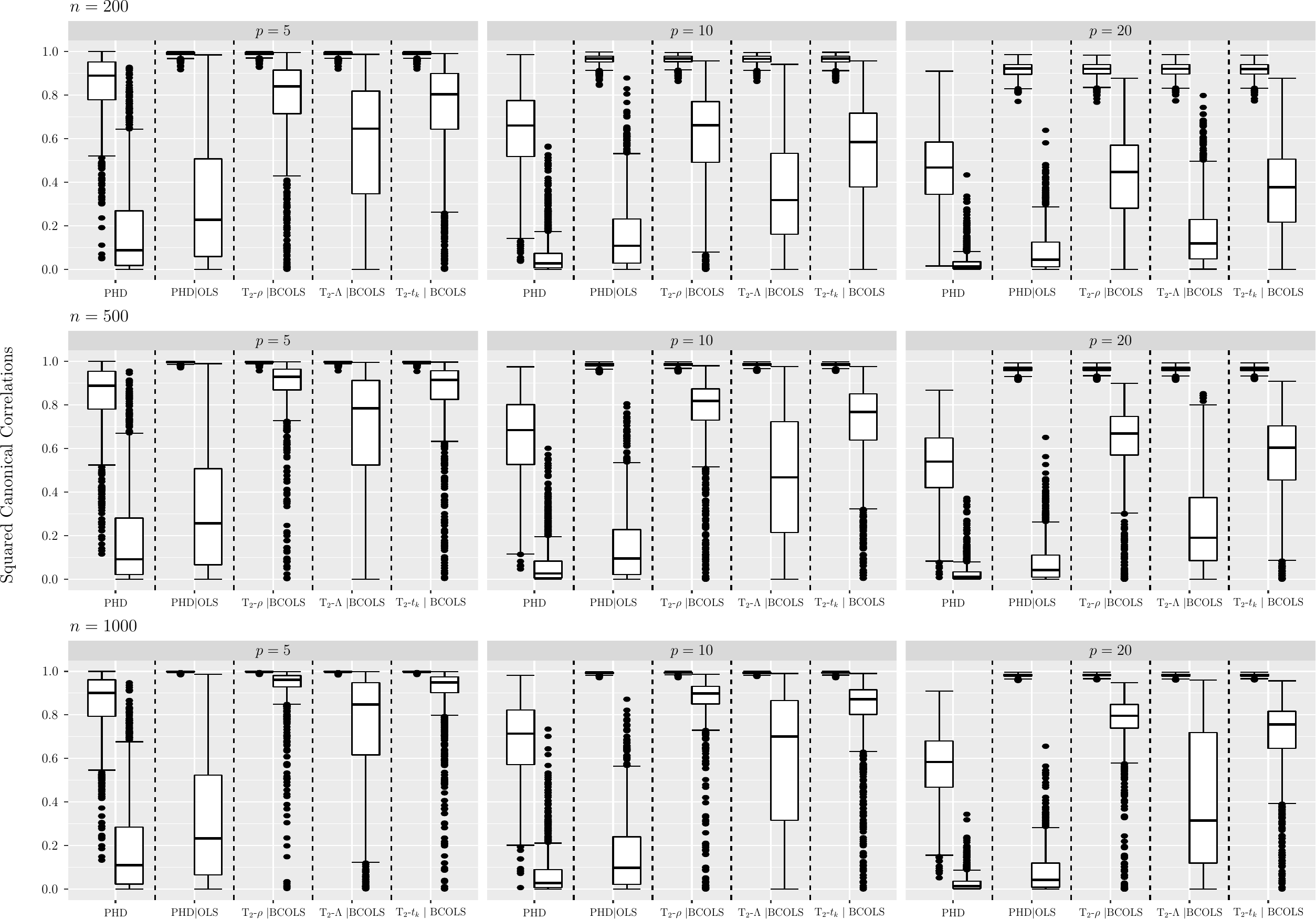}
    \caption{Boxplots of the squared canonical correlations of the PHD, PHD$|$OLS, $T_2$-PHD$_{\rho}$ $|$ BC-OLS, $T_2$-PHD$_{\Lambda}$ $|$ BC-OLS and $T_2$-PHD$_{t_k}$ $|$ BC-OLS methods (denoted by $T_2$-$\rho$ $|$ BCOLS, $T_2$-$\Lambda$ $|$ BCOLS and $T_2$-$t_k$ $|$ BCOLS respectively, for simplicity) across 1000 simulated runs from Model \ref{SinCubeModk2}, for three choices of $n$ (200, 500 and 1000) and three values of $p$ (5, 10 and 20). }
    \label{fig:SinCubeBox}
\end{sidewaysfigure}
}

The boxplots of the canonical correlations for Model \ref{SinCubeModk2}, in Figure \ref{fig:SinCubeBox}, show that PHD$|$OLS can greatly improve one of the directions, when dimensionality is low, but completely fails to estimate the other direction. However, we obtain significant improvements by the $T_2$-PHD$_{\rho}$ $|$BC-OLS and $T_2$-PHD$_{t_k}$ $|$BC-OLS methods, where the estimator variability shown in Table \ref{tab:SinCubeCors} is mainly attributed to one of the directions of the model. 

Finally, Table \ref{tab:sinCubeOptVals} shows the counts of the optimal values chosen by each iteration, for Model \ref{SinCubeModk2}, across the simulated runs. The interpretation of this table is very similar to the previous example with the difference of the BC-OLS iteration which shows very consistent choices across the different values of $p$. Note also that even though the $t_k$  and $\rho$ criteria choose different parameter values as optimal, the $T_2$-PHD$_{t_k}$ method still provides similar results to $T_2$-PHD$_{\rho}$.  
\begin{table}
\centering
\begin{adjustbox}{width = 0.83\textwidth, height = 0.125\textheight, keepaspectratio}
\begin{tabular}{r||r|rrr||r|rrr||r|rrr}
\toprule
 $p$ & \multicolumn{4}{c||}{5} & \multicolumn{4}{c||}{10} &  \multicolumn{4}{c}{20} \\
  \midrule
Criterion & $\rho_{ols}$ & \multicolumn{1}{c}{$\rho$} & \multicolumn{1}{c}{$\Lambda$} & \multicolumn{1}{c||}{$t_k$} & $\rho_{ols}$ & \multicolumn{1}{c}{$\rho$} & \multicolumn{1}{c}{$\Lambda$} & \multicolumn{1}{c||}{$t_k$} & $\rho_{ols}$ & \multicolumn{1}{c}{$\rho$} & \multicolumn{1}{c}{$\Lambda$} & \multicolumn{1}{c}{$t_k$} \\
  \midrule
  $\omega$ & $\omega_1$ & $\omega_2$ & $\omega_2$ & $\omega_2$ & $\omega_1$ & $\omega_2$ & $\omega_2$ & $\omega_2$ & $\omega_1$ & $\omega_2$ & $\omega_2$ & $\omega_2$ \\
  \toprule
$[-2,-1.6]$ &   0 &  17 & 160 &   0 &   0 &  16 & 302 &   0 &   0 &  17 & 471 &   0 \\ 
 $(-1.6,-1.2]$ &   0 &   8 & 103 &   0 &   0 &   8 & 115 &   0 &   0 &   2 & 144 &   0 \\ 
  $(-1.2,-0.8]$ &   0 &   5 & 129 &   0 &   0 &   5 &  96 &   0 &   0 &   3 &  80 &   0 \\ 
  $(-0.8,-0.4]$ &   2 &   9 & 126 &   0 &   0 &   9 &  76 &   0 &   0 &   8 &  37 &   0 \\ 
  $(-0.4,0]$ &   5 &  31 &  77 &   0 &   2 &  18 &  58 &   0 &   0 &  15 &  36 &   0 \\ 
 $(0,0.4]$ &  67 &  79 &  59 &   0 &  47 &  74 &  35 &   0 &  17 &  38 &  25 &   0 \\ 
  $(0.4,0.8]$ & 242 & 175 &  47 &   0 & 264 & 201 &  25 &   0 & 248 & 195 &  15 &   0 \\ 
  $(0.8,1.2]$ & 314 & 263 &  27 &   0 & 369 & 288 &  32 &   0 & 427 & 304 &  13 &   0 \\ 
  $(1.2,1.6]$ & 205 & 178 &  21 &   1 & 202 & 181 &  12 &   0 & 213 & 221 &   4 &   0 \\ 
  $(1.6,2]$ & 165 & 235 & 251 & 999 & 116 & 200 & 249 & 1000 &  95 & 197 & 175 & 1000 \\ 
   \bottomrule
\end{tabular}
\end{adjustbox}
\caption{Counts of Optimal parameter values of the $T_2$-PHD$_{\rho}$ $|$BC-OLS, $T_2$-PHD$_{\Lambda}$ $|$BC-OLS and $T_2$-PHD$_{t_k}$ $|$BC-OLS methods, for Model \ref{SinCubeModk2}, where $\omega_1$ is the optimal parameter value given by the BC-OLS method using the minimum influence criterion, denoted by $\rho_{ols}$, and $\omega_2$ the optimal parameter values of the $T_2$-PHD methods given by the three different criteria ($\rho$, $\Lambda$ and $t_k$), for $n=500$ and $p = 5, 10$ and 20.}
\label{tab:sinCubeOptVals}
\end{table}

\section{Example}\label{sec:RealData}

We consider the `bigmac' dataset taken from \cite{Rudolf}. The dataset contains the average values of 10 economic indicators in 1991 for 45 cities around the world. All prices are in US dollars, using currency conversion at the time of publication. We let the response be \textit{bigmac}, which is the minimum labor required to buy a Big Mac and fries from MacDonalds in each city. Information of the 9 predictors are included in Table \ref{tab:BigMacInfo}. 

For this data we compare the performance of the OLS and BC-OLS methods, for $K=1$. PHD, $T_1$-PHD and $T_2$-PHD were also considered but did not provide informative results. A second direction from the iterative OTDR methods also provided no additional information about the relationship between the \textit{bigmac} and the 9 economic indicators. 

\begin{figure}[h!]
    \centering
    \includegraphics[width = \textwidth, page = 1]{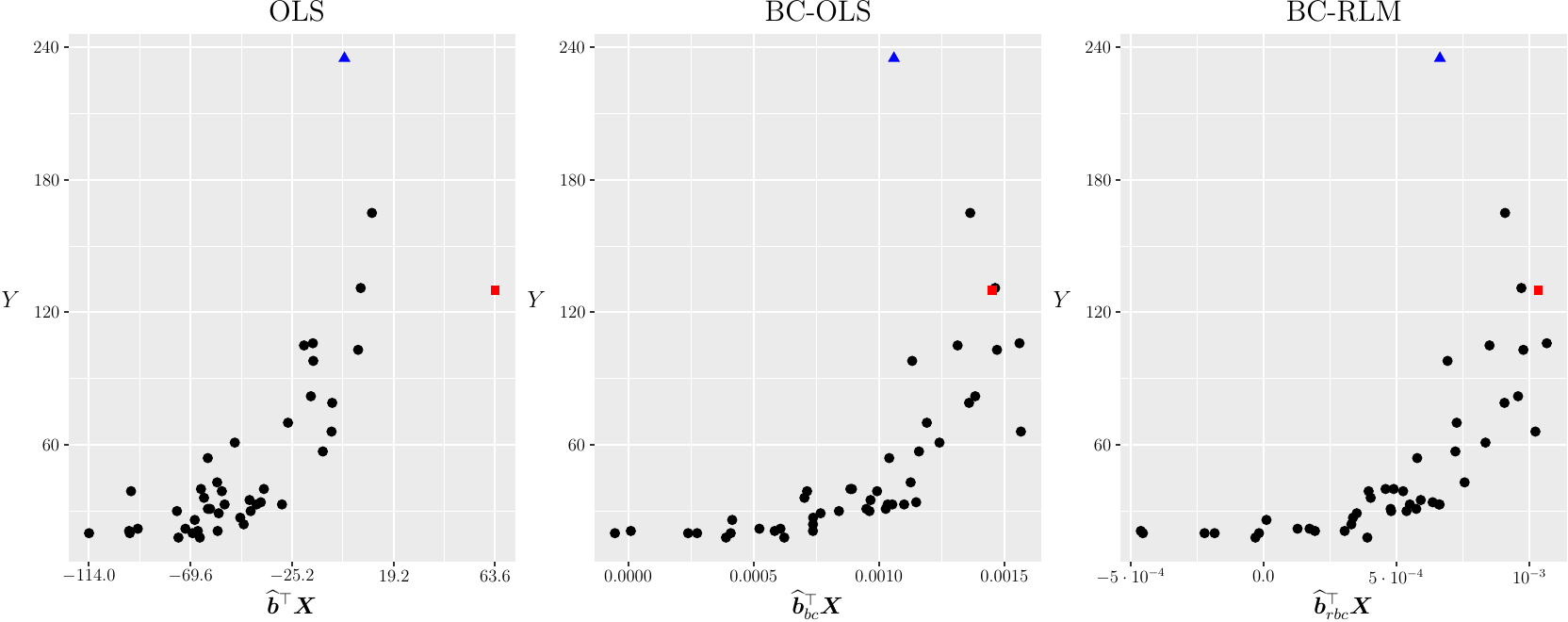}
    \caption{Caption}
    \label{fig: ESSPsBigMac1}
\end{figure}

Keep in mind, that it has previously been shown that removing outliers from the estimation of the OLS slope vector can provide improved ESSP's \citep{Olive2004}. Also, \cite{Prendergast2008} showed that trimming influential observations from the estimation but including them in the visualisation can have great benefits. Inspired by these, we used the influence measure given in \eqref{eq:rhoi} and found two influential observations in the data. By examining their behaviour, when removed from the data, we determined the presence of an outlier and a highly influential observation in the `bigmac' data. In the ESSP given by OLS in Figure \ref{fig: ESSPsBigMac1}, the outlier is the observation at the very top whereas the influential observation is the one at the far right of the plot, shown by a triangle and a square respectively.

\begin{table}[h]
    \centering
    \begin{tabular}{c|c}
\toprule
Name & Info \\ 
\midrule
Bread  &  Minimum labor to buy 1 kg bread \\
BusFare & Lowest cost of 10k public transit \\
EngSal & Electrical engineer annual salary, 1000s \\
EngTax &  Tax rate paid by engineers \\
Service & Annual cost of 19 services \\
TeachSal &  Primary teacher salary, 1000s \\
TeachTax &  Tax rate paid by primary teachers \\
VacDays &  Average days vacation per year\\
WorkHrs &  Average hours worked per year  \\
\bottomrule
    \end{tabular}
    \caption{Information on the predictor variables in the bigmac data.}
    \label{tab:BigMacInfo}
\end{table}

We also decided to perform the proposed method using robust linear regression to avoid removing any observations.   We denote this as BC-RLM and used the \verb rlm  function in R, which uses M-estimators that are less sensitive to outliers \citep[see, for example,][]{Huber1981}. Furthermore, note that influential observations when considering $Y$ might not be influential when considering the optimally transformed $Y$. The same can be true for outliers. 

The ESSPs in Figure \ref{fig: ESSPsBigMac1} indicate that the BC-OLS method shows a clearer relationship, similar to exponential growth, between $Y$ and $\widehat{\bm{b}}_{bc}^\top \X$ than the default OLS method. 
Finally, the ESSP given by BC-RLM shows an even sharper view of the relationship between $Y$ and $\widehat{\bm{b}}_{rbc}^\top \X$.

Note here that the BC-OLS and BC-RLM methods chose the same optimal parameter value, $\omega = -2$, for the transformation. The optimal response transformation that provides the improved \edr estimates, $\widehat{\bm{b}}_{bc}$, $\widehat{\bm{b}}_{rbc}$, for the data is, $\dfrac{\textit{bigmac}^{-2} - 1}{-2}$.

\section{Discussion and further work}\label{sec:Conclusion}

In this article, we demonstrated how response transformations can greatly improve the estimation of the \edr directions in dimension reduction with OLS and PHD. We have provided an automated method that searches for the optimal transformation for a given model while using the influence measure \citep{Smith10} as a criterion to find the optimal parameter value of the transformation. Alternative criteria for choosing the optimal transformation have also been provided for time-efficiency in practice, which were shown to be able to perform almost as good as the minimum influence criterion. An iterative approach of this method was also provided to further improve estimation for the second or more directions. Simulated comparisons  and a real-world example highlighted the success of the methods proposed and showed that we can achieve improved visualizations of the relationship between the response and predictor variables. 

This method can be extended further by considering more transformations and using them to improve other dimension reduction techniques which then allows for more iterative dimension reduction combination methods. 

\bibliographystyle{authordate4}
\bibliography{references}  

\end{document}